 \documentclass[acmsmall]{acmart}

\usepackage{amsmath}
\usepackage{graphicx}
\usepackage{algorithm}
\usepackage{algorithmicx}
\usepackage{algpseudocode}
\usepackage{subcaption}

\usepackage{booktabs}
\usepackage{tabularx}
\usepackage{hyperref}

\usepackage{booktabs}
\usepackage{longtable}
\usepackage{multirow}

\AtBeginDocument{%
  }

\setcopyright{acmlicensed}
\copyrightyear{2025}
\acmYear{2025}
\acmDOI{XXXXXXX.XXXXXXX}

\acmJournal{JACM}
\acmVolume{37}
\acmNumber{4}
\acmArticle{111}
\acmMonth{8}

\begin{document}


\title[Cybersecurity Competitions in K-12]{Beyond the Flag: A Framework for Integrating Cybersecurity Competitions into K-12 Education for Cognitive Apprenticeship and Ethical Skill Development}

 \author{Tran Duc Le}
 \affiliation{%
   \institution{University of Wisconsin–Stout}
   \city{Menomonie}
   \state{WI}
   \country{USA}}
 \email{let@uwstout.edu} 

 \author{Truong Duy Dinh}
 \affiliation{%
   \institution{Posts and Telecommunications Institute of Technology}
  \city{Hanoi}
   \country{Vietnam}}
 \email{duydt@ptit.edu.vn}

 \author{Phuc Hao Do}
 \affiliation{%
   \institution{Danang Architecture University}
   \city{Da Nang}
   \country{Vietnam}}
 \email{haodp@dau.edu.vn}

 \author{Van Dai Pham}
 \affiliation{%
   \institution{Swinburne Vietnam, FPT University}
   \city{Hanoi}
   \country{Vietnam}}
 \email{daipv11@fe.edu.vn} 

 \author{Nam Son Nguyen}
 \affiliation{%
   \institution{Hewlett Packard Enterprise}
   \city{Bloomington}
   \state{MN}
   \country{USA}}
 \email{nam.nguyen@hpe.com} 

 \renewcommand{\shortauthors}{Tran Duc Le et al.}

\begin{abstract}
Capture the Flag (CTF) competitions are powerful pedagogical tools for addressing the global cybersecurity workforce gap, yet their effective K-12 implementation is often undermined by significant barriers, including educator preparedness gaps and equity concerns. This paper addresses these challenges by proposing the Ethical-Cognitive Apprenticeship in Cybersecurity (ECAC) framework, a new model derived from a systematic Framework Synthesis of existing literature and empirical evidence. ECAC systematically integrates cognitive apprenticeship theory with embedded ethical development across five phases: (1) Foundational Modeling, (2) Scaffolding the Arena, (3) Coaching and Articulation, (4) Ethical Dilemma Injections, and (5) Reflective Exploration. The framework provides a "low floor, high ceiling" learning pathway designed to broaden participation among diverse student groups, including underrepresented minorities and women, while fostering deep, transferable skills. By reframing the educator’s role as a lead learner," ECAC also offers a sustainable solution to the teacher expertise gap. Ultimately, this framework provides a practical roadmap for transforming CTFs from standalone competitions into integral learning experiences that cultivate a more skilled, ethical, and diverse generation of cybersecurity professionals.
\end{abstract}



\received{xx xxx xxxx}
\received[revised]{xx xxx 2025}
\received[accepted]{xx xxx 2025}

\maketitle

\section{Introduction}
\label{sec1}
A persistent and widening cybersecurity workforce gap presents a critical challenge to national security and economic stability. Recent analyses document a global shortage of millions of skilled professionals required to defend essential digital infrastructure, with projections indicating that this talent deficit poses a direct threat to global economies \cite{Ibrahim2024}. The escalating sophistication and frequency of cyber threats have elevated cybersecurity from a specialized technical field to a national security imperative, demanding an urgent and innovative strategy for cultivating a sustainable talent pipeline. In addition, There is broad consensus that cultivating interest and skills in cybersecurity during the K-12 years can help address the critical workforce shortage in this field while also promoting safer online behaviors among youth \cite{Ford2017,Jin2018}.

In response, national strategies increasingly recognize that the foundation for a resilient cybersecurity workforce must be established long before the collegiate level \cite{Dill,Blai2021}. Studies in the EU indicate that cybersecurity topics are frequently missing from informatics programs in high schools \cite{Blai2024}. Similarly, at the primary level, traditional teaching methods have proven ineffective in engaging students with cybersecurity concepts \cite{Blai2024}. Consequently, K-12 education is now understood as the critical stage for instilling foundational interest and digital literacy to guide a diverse student population toward technology-focused careers \cite{Ford2017,Dawson2022}. Deferring the introduction of core cybersecurity concepts until university is insufficient to capture the broad pool of talent required to meet accelerating demand.

Addressing the workforce imperative, however, requires overcoming a significant pedagogical gap. Traditional educational models, often reliant on passive instruction through lectures and textbooks, are ill-suited to convey the dynamic, hands-on nature of the cybersecurity field. Such methods frequently prove ineffective at fostering genuine student interest and fail to develop the durable, practical problem-solving skills that are critical for professional success, leading to lower skill retention compared to active learning approaches \cite{Ford2017,Qusa2021}. Consequently, many students acquire abstract theoretical knowledge but cannot apply it to the complex, real-world scenarios that define cybersecurity practice, creating a disconnect between academic instruction and practical competency \cite{Casey2023, Li}. This gap necessitates a shift toward innovative and experiential learning strategies capable of bridging theory and practice for a K-12 audience.
In this context, Capture the Flag (CTF) competitions emerge as a powerful pedagogical model. CTFs are gamified challenges where participants solve security-related puzzles to find digital "flags," transforming abstract principles of cryptography, digital forensics, or network defense into tangible, competitive experiences \cite{White2010}. This approach aligns with modern educational theories emphasizing active learning and constructivist pedagogy \cite{OConnor2020}, providing authentic, simulated environments where students can legally explore the field’s core tenets \cite{Jin2018}.

The existing literature confirms that CTFs are effective pedagogical tools. Studies consistently show they enhance student engagement and motivation, foster career interest, and improve technical competencies and higher-order thinking skills such as persistence and critical thinking \cite{Ibrahim2020, Gough2024, Purbo2024, VidenovikFili2024}. The educational power of CTFs is rooted in three key pedagogical theories that make them potent tools for K-12 cybersecurity education:
\begin{itemize}
\item \textbf{Gamification:} CTFs exemplify gamified learning through game-design elements such as points, leaderboards, narratives, and immediate feedback that enhance motivation and engagement \cite{Deterding2011}. The point systems and leaderboards provide clear measures of progress and foster healthy competition, while narrative elements frame challenges as exciting missions rather than dry academic exercises \cite{Casey2023}. This approach transforms dense technical material into intrinsically motivating experiences that capture student interest.
\item \textbf{Problem-Based Learning (PBL):} CTFs function as student-centered environments where learning occurs through solving authentic, open-ended problems \cite{Yew2016}. Instead of being lectured on topics, students are presented with challenges that compel them to actively discover, understand, and apply concepts. This process requires students to identify their own knowledge gaps and engage in self-directed learning, leading to deeper understanding and better knowledge retention \cite{Purbo2024,VidenovikVold2024}.
\item \textbf{Cognitive Apprenticeship:} CTFs create environments that facilitate cognitive apprenticeship, where learners acquire skills through interaction with expert thinking processes made visible \cite{collins1991cognitive}. In CTFs, the "expert" is embodied by challenge design and the community of practice. The challenges serve as modeling, demonstrating expert approaches to vulnerability analysis \cite{Toccafondi2024}. Coaching and scaffolding occur through hint systems and tiered difficulty levels. Team-based CTFs require articulation and reflection as students collaborate and compare methods with expert solutions \cite{VidenovikFili2024,Cherinka2018}. Finally, CTFs encourage exploration in safe, sandboxed environments \cite{Toccafondi2024}.
\end{itemize}

Recent empirical research has demonstrated the effectiveness of innovative approaches that address traditional CTF limitations. Rathore and Griffith showed that collaborative scavenger hunt activities achieved 77.4\% student satisfaction rates \cite{Rathore}. Gough et al. and Namukasa et al. provided compelling evidence that collaborative, scaffolded approaches significantly increased engagement among underrepresented minorities and women, with females showing significantly higher intent to pursue cybersecurity careers after participation \cite{Gough2024, Namukasa2024}.

The critical question for the field is therefore no longer if CTFs are effective, but how to implement them effectively, equitably, and sustainably across diverse K-12 settings. There remains a significant gap in the literature for a comprehensive, evidence-based framework that moves beyond a simple list of best practices to guide educators, administrators, and policymakers.

To address the limitations of current cybersecurity competition models and foster more profound learning, the integration of established pedagogical theories is essential. Cognitive Apprenticeship (CA) emerges as a particularly potent model for teaching complex cognitive skills, such as those required in cybersecurity, within authentic, problem-solving contexts \cite{Dawson2018,Heverin2024}. CA focuses on making expert thinking processes visible and learnable through methods like modeling, coaching, and scaffolding, which are highly relevant for demystifying intricate cybersecurity concepts for K-12 learners \cite{Lamond2022}.

Simultaneously, the development of ethical reasoning is a non-negotiable component of comprehensive cybersecurity education, especially for K-12 students who are in critical stages of moral development. The aim to cultivate "conscious and responsible digital citizens from early ages" necessitates that ethical considerations are not merely an addendum but are deeply interwoven with technical skill acquisition.

This paper aims to fill this gap by developing and validating the Ethical-Cognitive Apprenticeship in Cybersecurity (ECAC) framework. We argue that for CTFs to realize their full potential, they must be intentionally structured as sites for cognitive apprenticeship, where students learn to think like experts, and where ethical skill development is treated as a core, non-negotiable component. To build this argument, the paper is structured as follows: Section 2 reviews the theoretical foundations of this approach. Section 3 details the Framework Synthesis methodology, while Section 4 presents the findings on CTF impacts and implementation barriers. Section 5 then introduces the proposed five-phase ECAC framework in detail. Following this, Section 6 provides concrete examples of the framework's staged implementation across K-12 grade levels, and Section 7 validates the framework through a multi-method analysis. Finally, Section 8 discusses the framework’s implications for stakeholders, Section 9 addresses its limitations and future research, and Section 10 provides a conclusion that summarizes the paper's contributions.

\section{Theoretical Underpinnings}
\label{sec2}
\subsection{Capture the Flag as Learning Environments}
\label{sec2.1}

To understand the educational impact of CTF competitions, it is essential to define their structure and place in the K-12 context. At its core, a CTF is an exercise in applied cybersecurity problem-solving where participants locate a “flag” – a specific string of text – hidden within a vulnerable system or digital artifact \cite{White2010}. Originally emerging from hacker conferences like DEF CON (\url{https://defcon.org} [Access date: June 15, 2025]), CTFs have been widely adapted for educational purposes \cite{McDaniel2016}. The challenges are designed to mirror real-world tasks in categories such as web exploitation, cryptography, digital forensics, and reverse engineering \cite{Rathore,Balon2023}. In K-12 education, the most prevalent format is \textbf{Jeopardy-style}, where challenges are arranged in a grid of categories with point values corresponding to difficulty. This structure is favored for its accessibility and scalability, allowing students to self-direct their learning and build confidence by starting with easier problems \cite{Rathore}. A more dynamic but less common format at the K-12 level is \textbf{Attack-Defense (A/D)}, where teams defend their own services while simultaneously attempting to exploit opponents'. While the A/D format closely simulates real-world cyber conflict, its technical complexity and resource requirements make it less suitable for introductory school contexts \cite{Rathore}.

The cognitive apprenticeship elements naturally embedded in CTF environments make them particularly powerful for developing expert-like thinking. The goal extends beyond knowledge transfer to enculturating students into authentic cybersecurity practices and problem-solving mindsets, elevating educational outcomes from simple skill acquisition to the development of expert-like thinking \cite{Toccafondi2024}.

\subsection{CTFs as a Site for Cognitive Apprenticeship}
\label{sec2.2}
Beyond gamification and PBL, we argue that the most potent aspect of CTFs is their ability to function as an environment for \textbf{cognitive apprenticeship}, a model where novices learn to think like experts by observing and engaging in the authentic practices of a community \cite{Collins2018}. Unlike traditional apprenticeships focused on physical crafts, CA targets cognitive and metacognitive skills, making expert thought processes accessible to novices. This is achieved through a set of structured teaching methods designed to guide learners from observation to independent practice.

In a well-designed CTF ecosystem, expert thinking is made visible through several core methods that parallel the stages of cognitive apprenticeship \cite{Matsuo2024}:
\begin{itemize}
\item \textbf{Modeling:} Experts (teachers, mentors, or advanced peers) demonstrate the target skills and, crucially, their underlying thought processes. In cybersecurity, this could involve demonstrating how to approach a CTF challenge, troubleshoot a network issue, or reason through an ethical dilemma, making the implicit strategies of experts explicit.
\item \textbf{Coaching:} This involves observing students as they undertake tasks (e.g., attempting CTF challenges) and offering timely, individualized guidance. This guidance can take the form of hints, feedback, leading questions, reminders, or the introduction of new, related tasks to extend their learning.
\item \textbf{Scaffolding:} Scaffolding refers to the provision of temporary support structures that enable students to perform tasks that would otherwise be beyond their current capabilities. In K-12 cybersecurity, scaffolds could include simplified CTF challenges, pre-configured virtual environments, "\textit{unplugged}" activities that explain complex concepts without technology, conceptual guides, or partial solutions. These supports are gradually withdrawn (faded) as students develop proficiency.
\item \textbf{Articulation:} This method encourages students to verbalize their knowledge, reasoning, and problem-solving strategies. This can occur through explaining how they solved a CTF challenge, justifying an ethical decision, or teaching a concept to a peer. Articulation helps students clarify their understanding and makes their thinking visible to both themselves and their instructors.
\item Reflection: Reflection prompts students to critically evaluate their own learning processes and performance. This often involves comparing their problem-solving approaches and ethical reasoning with those of experts, peers, or established best practices. Post-CTF write-ups or group discussions can facilitate reflection on strategies, successes, failures, and ethical choices made during the competition.
\item \textbf{Exploration:} This method encourages learner autonomy by setting general goals and allowing students to define and pursue their own problems or areas of interest. In a cybersecurity context, this could involve students independently researching new vulnerabilities, experimenting with different tools (in safe environments), or even designing their own simple CTF challenges for peers.
\end{itemize}
Several studies in computing education suggest using apprenticeship models for complex skill acquisition. For example, Gough et al. implicitly adopted a cognitive apprenticeship stance by employing “\textit{direct mentoring as [a] supportive strategy}” in their collaborative IoT hacking activities, literally having instructors and near-peer mentors guide students through the problem-solving process \cite{Gough2024}. They concluded that scaffolding and mentorship were key to engaging historically marginalized groups who might have been left behind in pure competition. Similarly, in the \textit{GenCyber} context, camp instructors have noted that walking through the first challenge together as a group (an expert-led demonstration) helped set students up for success when tackling subsequent challenges individually \cite{McDaniel2016}. This is a clear modeling phase. During the competition, instructors and volunteer experts often circulate to give hints or reframe problems – functioning as just-in-time scaffolding and coaching \cite{McDaniel2016}. Then, as students gain confidence, they tackle more on their own, representing the fading of support.

CA also emphasizes the \textbf{social nature of learning}, encouraging interaction between novices and experts and among peers. This resonates with the peer learning and teamwork benefits mentioned earlier \cite{VidenovikFili2024}. In a classroom CTF integration, a teacher would not simply assign problems and wait for students to finish; rather, the teacher actively coaches students, perhaps thinking aloud to demonstrate how to approach a challenge (modeling analytical reasoning), providing hints like a tool or concept to try (scaffolding), and encouraging students to explain their solution strategy to the class (articulation and reflection). Videnovik et al. actually applied a related strategy by pairing older students as peer mentors to younger ones during game-based cybersecurity lessons, effectively creating an environment where students could observe more experienced peers and then practice with guidance \cite{VidenovikFili2024}. The result was improved learning outcomes and confidence for both mentors and mentees.

The CA model is highly pertinent to cybersecurity education because it deals with \textbf{tacit knowledge and strategies} that are hard to learn through textbooks alone (e.g., how to troubleshoot why an exploit failed, or how to systematically search for vulnerabilities). These are best learned by “\textit{learning by doing}” with an expert’s guidance. In fact, Qusa \& Tarazi cite that people generally retain only 20\% of what they hear or read but up to 90\% of what they actively do \cite{Qusa2021}. This supports embedding active, authentic tasks (like CTF challenges) in instruction. They designed a \textit{Cyber-Hero} framework for high school cybersecurity awareness which gamified the process of learning password security by having students practice creating strong passwords in a game scenario \cite{Qusa2021}. The framework involved iterative practice with feedback, mirroring the idea of apprenticeship where a skill is honed through repeated attempts under a master’s supervision. Early results from that work showed incremental progress in student capabilities, validating the “\textit{practice with guidance}” approach.

Finally, applying CA effectively in K-12 cybersecurity education requires careful adaptation of these methods to suit the developmental stages of learners and the specific nature of cybersecurity tasks. CTFs, with their inherent problem-solving nature, can provide the authentic, situated learning contexts that CA emphasizes \cite{Bowen2022}. The structured yet engaging environment of a well-designed CTF allows for the systematic application of CA principles, making the often-invisible cognitive processes of cybersecurity experts more tangible for K-12 students \cite{AhmadHaziqAshrofieHanafi2021,Nagare2025}.

A key aspect of this adaptation is the principle of progressive complexity and diversity of tasks. Cybersecurity concepts and CTF challenges must be introduced in a way that builds incrementally on prior knowledge, starting with foundational ideas and gradually moving towards more complex scenarios. For instance, "modeling" for younger K-5 students might involve interactive storytelling or role-playing basic cyber hygiene practices, like those seen in playful educational games. For older high school students, modeling could take the form of live demonstrations by teachers or industry professionals using ethical hacking tools in a controlled CTF environment \cite{Casey2023}. Without such age-specific adaptation, a generic application of CA could lead to cognitive overload for younger learners or insufficient challenge for older students, thereby undermining the framework's effectiveness. The ECAC's proposed staged approach is therefore essential for tailoring CA methods appropriately across the diverse K-12 landscape.

\subsection{The Imperative of Ethics in Cybersecurity Education}
\label{sec2.3}
The development of ethical reasoning is paramount in cybersecurity education, particularly at the K-12 level where students are forming their fundamental values and understanding of societal norms.

\subsubsection{The Imperative of Ethics in Cybersecurity Education}
\label{sec2.3.1}

The "dual-use" nature of cybersecurity knowledge – where skills can be used for both defensive and offensive purposes – makes a strong ethical foundation indispensable \cite{Macnish2020}. Education must go beyond technical training to instill a sense of responsibility, an understanding of the potential societal impact of cyber actions, and a commitment to using skills ethically. This includes promoting "white hat" values and responsible digital citizenship from an early age. The "Ethics or Death" module described in one program, which starts with an ethical foundation, underscores this imperative \cite{Gough2024}.

\subsubsection{Developmental Considerations for Ethical Learning in K-12}
\label{sec2.3.2}

Cybersecurity, as a discipline, exists at the intersection of technology and human behavior, raising issues of privacy, security, and ethics. Therefore, purely technical training risks creating skilled individuals who might lack the judgment to apply their skills responsibly. Several educational initiatives and research efforts emphasize weaving \textbf{ethics and digital citizenship }into cybersecurity learning. Digital citizenship refers to responsible and appropriate use of technology – including understanding online safety, privacy, intellectual property, cyberbullying, etc. Many K-12 curricula already include digital citizenship modules (often in computing or advisory courses), and cybersecurity education can complement and extend these topics \cite{VidenovikFili2024,Chattopadhyay2024}. For example, basic positive digital behaviors like creating strong passwords, not sharing personal info, and being wary of phishing are foundational cybersecurity lessons usually taught as part of digital citizenship programs \cite{Kuchar2023}. The challenge and opportunity are to integrate those lessons with hands-on security activities.

Casey et al. provide a strong example of combining technical and ethical learning \cite{Casey2023}. In their \textit{Cyber Sleuth Science Lab} for girls, students not only learned forensic techniques but also explored “\textit{complex social issues related to technology}” and engaged in discussions and mock trials about evidence and ethics. Importantly, they reported that their gender-inclusive strategies (which included highlighting real-world relevance and community impact) did “\textit{engage girls without being harmful to boys’ engagement}”. This reinforces that integrating ethics and social context can increase appeal to a broader student base, while still benefiting all. After participating, girls showed a greater increase than boys in interest in pursuing related careers, indicating that the ethical/social dimension might have particularly resonated with them \cite{Casey2023}.

In the context of CTF integration, ethical skill development means a few things: teaching students about the \textbf{ethical guidelines of cybersecurity} (for instance, the concept of responsible disclosure, the illegality of unauthorized system access, respecting privacy even as they learn to defeat security mechanisms) and fostering a sense of \textbf{purpose and ethics }(encouraging students to see themselves as guardians rather than attackers, emphasizing defense and protection). A better approach, according to experts, is to incorporate discussion and reflection around each challenge – for instance, after a challenge on cracking passwords, a teacher might lead a conversation: how could a malicious actor misuse this skill? How can we defend against it? What are the legal consequences of doing this outside a safe environment? This reflection ties back to cognitive apprenticeship’s articulation phase and simultaneously builds ethical reasoning. As an example, the NICE framework (a guiding framework for cybersecurity education and workforce) \cite{Petersen2020} and other frameworks underscore the importance of ethics \cite{Crabb2024}. GenCyber’s first fundamental principle is actually “Ethics – understanding ethical behavior in cyberspace” (phrased in some documents as “All GenCyber participants are expected to practice ethical behavior in cyberspace”) \cite{Chattopadhyay2021}. This principle was established to ensure that as we excite students about hacking, we simultaneously impart a moral compass. Thus, our framework makes ethics and digital citizenship one of its pillars to ensure these crucial aspects are not an afterthought but an integral part of the learning process.

In conclusion, the theoretical and empirical insights from prior work suggest that a successful K-12 cybersecurity education approach will treat learning as a \textbf{mentored practice} (cognitive apprenticeship) and cybersecurity competency as including \textbf{ethical understanding} (not just technical skill). These inform the design of our framework’s components.

\section{Methodology}
\label{sec3}

\subsection{Research Approach: Framework Synthesis}
\label{sec3.1}
To develop a practical, evidence-based model for K-12 CTF integration, this study employed a \textbf{Framework Synthesis} methodology. This approach is a form of qualitative synthesis specifically designed for policy- and practice-oriented questions \cite{Ritchie}. Unlike a purely inductive thematic analysis where themes emerge entirely from the data, this approach uses a set of pre-defined research questions to guide a largely deductive process of organizing and analyzing findings from diverse sources. This method was chosen because it directly aligns with the paper's objective: to synthesize evidence related to the specific, pre-identified domains of CTF impacts, implementation barriers, and best practices to build a usable framework for educators.

\subsection{Literature Search and Selection}
\label{sec3.2}
A systematic literature search was conducted using the SCOPUS database to build the evidence base. The search was designed to be both comprehensive and precise, targeting peer-reviewed literature at the intersection of gamified learning, cybersecurity, and K-12 education. The following query was used:

\textit{TITLE-ABS(("Capture the Flag" OR "CTF" OR "gamification" OR "game-based learning" OR "competition-based learning") AND ("cybersecurity" OR "cyber security") AND ("K-12" OR "K12" OR "high school" OR "secondary school" OR "middle school" OR "school" OR "youth"))}

This search yielded 48 results, which were then manually filtered according to strict inclusion criteria. Included sources were peer-reviewed academic publications with a primary focus on CTFs or gamified cybersecurity education in a K-12 or analogous early-undergraduate context. Sources were excluded if they focused exclusively on professional-level competitions, were purely technical descriptions without pedagogical analysis, were not available in English, or were inaccessible. This rigorous filtering process resulted in the final corpus of 25 papers (See \ref{app1}.) used for this synthesis.

\subsection{Data Extraction and Analysis}
\label{sec3.3}

A structured data extraction protocol was developed to systematically extract relevant information from selected papers. The extraction was guided by four key Research Questions (RQs) aligning with core manuscript objectives:
\begin{itemize}
\item \textbf{RQ1:} \textit{What are the impacts of CTFs on students?}
\item \textbf{RQ2:} \textit{What are the barriers to implementing CTFs in K-12 settings?}
\item \textbf{RQ3:} \textit{What best practices or strategies for designing and running successful K-12 CTF programs are reported?}
\item \textbf{RQ4:} \textit{How are ethical considerations addressed in these CTF initiatives?}
\end{itemize}
For each paper, relevant text segments corresponding to these questions were extracted and logged into a data matrix. We used an inductive thematic analysis approach. We paid particular attention to evidence pertaining to: (a) benefits and outcomes of using cybersecurity competitions or games in education; (b) challenges or barriers reported; and (c) any suggested solutions or frameworks. As we reviewed, certain recurring themes emerged (for example, “need for teacher training” or “importance of scaffolding tasks”). The initial set of themes identified included: student engagement, skill acquisition, gender and diversity issues, teacher readiness, curriculum integration, ethical considerations, resource constraints, scaffolding/pedagogy, and partnerships. We then organized these into larger categories aligned with our research objectives.

The next step was theoretical integration: we mapped the themes onto relevant educational theories (cognitive apprenticeship, constructivism, etc.) to ensure our framework would have a strong conceptual foundation. For instance, themes of “scaffolding” and “modeling” were aligned with cognitive apprenticeship principles, while themes of “ethics” and “digital citizenship” were aligned with established goals in character education and computing ethics. We also looked at where multiple studies independently pointed to similar recommendations. For example, multiple sources highlighted scaffolded learning pathways as crucial \cite{VidenovikFili2024,McDaniel2016}, and numerous studies cited teacher professional development as a bottleneck \cite{Blai2024}. We considered such convergence as validation that those should be pillars in the framework.

No human subjects or experimental data were involved in this research. However, the methodology does rely on the credibility of the analyzed sources. To ensure robustness, we prioritized peer-reviewed academic sources and official educational frameworks, and we cross-referenced points across multiple sources. In writing the framework proposal, we cite specific studies to provide empirical grounding for each recommendation.

In essence, our methodology can be seen as a form of structured literature review driving conceptual model building. This approach is appropriate given the nascent state of K-12 cybersecurity education research – a fully data-driven derivation (e.g., meta-analysis) is impractical due to limited and heterogeneous studies, but a thoughtful synthesis can yield a novel framework that others can subsequently implement and empirically test.

It is worth noting that our framework has not yet been empirically tested as a whole (that is beyond the scope of this paper), but each component is supported by prior research findings. We will consider how the framework might be evaluated in future work.

The limitations of this methodology largely stem from the available literature: if certain biases exist in published studies (for example, most studies being from certain countries or focusing only on short-term interventions), those could influence our framework. We mitigated this by including a variety of sources (international studies, different educational contexts) and looking to general educational theory for guidance as well.

Having outlined how we derived our framework, we now transition to presenting the Findings from our literature analysis that directly inform each component. These findings serve as a bridge between the raw literature review and the concrete framework proposal that follows.

\section{Findings}
\label{sec4}

\subsection{Technical Skill Acquisition and Application}
\label{sec4.1}

One of the most consistently reported outcomes in the literature is that CTF-style activities provide an effective hands-on avenue for acquiring and applying technical cybersecurity skills. By immersing learners in problem-solving tasks that mirror real-world scenarios, CTFs effectively bridge the gap between abstract theory and practical competency \cite{Namukasa2024,Kikkas2020}. This practical "\textit{learning-by-doing}" is crucial, as studies note that hands-on practice facilitates knowledge retention and enables students to seamlessly apply their skills in authentic situations \cite{Lazarov2025}. The evidence shows that through these challenges, students gain proficiency in a wide array of core cybersecurity domains, including applied cryptography, steganography, digital forensics, penetration testing, and network security analysis \cite{Blai2024,Ibrahim2020,VidenovikFili2024,Rathore}.

The development of these skills can be remarkably advanced. For instance, participants learn to use fundamental command-line tools and identify common web vulnerabilities like SQL injection and Cross-Site Scripting (XSS) \cite{Ibrahim2020,McDaniel2016}. In a landmark analysis of the high school-focused \textit{picoCTF} (\url{https://picoctf.org} [Access date: June 15, 2025]), Chapman et al. found that participants successfully acquired skills typically taught at the university level, with many learning to forge HTTP cookies and a notable number executing complex return-oriented programming (ROP) exploits \cite{Chapman}. This tangible skill gain is validated across multiple studies. In one program, high school students with no prior security experience progressed to using tools like \textit{Wireshark} to capture live traffic \cite{McDaniel2016}, while another multi-session program saw students achieve an average success rate above 70\% on CTF tasks \cite{Lazarov2025}. Furthermore, self-reported data shows significant gains, with over 90\% of students in one camp series reporting a better understanding of cybersecurity concepts post-participation \cite{Ford2017}.

Beyond mastering specific tools, this applied context builds a more robust technical intuition, helping students \textit{"understand how things work under the hood.}" This deeper comprehension is described by some as understanding the "\textit{security implications of system intricacies}" \cite{Rathore}. This pedagogical benefit can be further enhanced by inverting the model and tasking students with designing their own challenges, a process that compels them to meticulously scrutinize underlying theory to create solvable puzzles \cite{Rathore}. Ultimately, the literature confirms that CTF participation provides an applied context that lectures alone cannot offer, leading to skills that are transferable to professional settings, as participants can "\textit{seamlessly use [new skills] in real-world situations}" \cite{Lazarov2025}.

\subsection{Engagement, Motivation, and Self-Efficacy}
\label{sec4.2}
Beyond technical skills, the literature provides overwhelming evidence of the affective impact of CTFs. The gamified design is a powerful catalyst for engagement, transforming cybersecurity learning into an exciting and intrinsically motivating pursuit. Competitive elements like points, leaderboards, and timers foster a stimulating atmosphere that encourages students to persist through difficulty and invest significant discretionary effort \cite{Gough2024,Purbo2024,Chattopadhyay2024}. This effect is quantitatively validated by studies such as Chapman et al., who noted that K-12 students voluntarily spent 11–12 hours on average actively solving challenges – a level of engagement rarely seen in traditional coursework \cite{Chapman}. Survey data further reflects this, with one study of a high school workshop finding that 92 of 96 respondents indicated moderate to great enjoyment in solving the CTF challenges \cite{Saddiqa2024}. This motivational pull can also inspire continued learning long after an event concludes \cite{Ford2017}.

Furthermore, this deep engagement is fundamental to building student confidence and self-efficacy. Many successful K-12 CTFs employ a scaffolded, "\textit{low floor, high ceiling}" design, which is critical to this process \cite{Purbo2024,VidenovikFili2024}. By providing introductory exercises that virtually all students can solve, programs allow participants to capture an initial flag, which provides immediate positive reinforcement and builds the confidence needed to tackle future obstacles \cite{McDaniel2016}. This process of building self-efficacy through incremental achievement has been identified as particularly critical for engaging and retaining students from underrepresented groups, who may enter with less confidence in their technical abilities \cite{Chattopadhyay2024,Ford2017}.

However, the literature also indicates that the competitive nature of CTFs must be carefully managed to ensure these positive outcomes are equitable. While many students thrive on competition, a subset can experience stress or fear of failure, and design choices like deducting points for hints can inadvertently deter help-seeking \cite{Lazarov2025}. Moreover, engagement is not always uniform across demographics. Research has highlighted potential gender disparities in the enjoyment of gamified formats, indicating that simply adding competition may not engage all student groups equally \cite{Qusa2021}. This finding underscores the need for thoughtful and inclusive design. On a positive note, tailored interventions show significant promise; for example, an all-girls STEM camp that used a cryptography scavenger hunt elicited "\textit{significant enthusiasm}" and spurred interest in cybersecurity careers \cite{Rathore}. Therefore, the literature suggests that while CTFs are potent motivational tools, their ultimate success hinges on a design that scaffolds difficulty, mitigates negative competitive pressures, and is consciously inclusive.

\subsection{Higher-Order Thinking and Collaborative Skills}
\label{sec4.3}

The literature consistently indicates that CTFs cultivate advanced cognitive skills, pushing students beyond rote memorization to engage in higher-order thinking. Unlike guided exercises, the open-ended nature of CTF challenges requires students to systematically deconstruct problems, test hypotheses, and adapt their strategies, fostering what is described as “\textit{investigative critical thinking}” \cite{Ford2017,Qusa2021,Purbo2024,Namukasa2024}. This process helps develop a “\textit{security mindset}” – an expert-like ability to analyze systems not just for their intended function but for their potential vulnerabilities \cite{Blai2024,Toccafondi2024}. This can even spark creativity and out-of-the-box thinking, as seen when students discover and exploit unadvertised vulnerabilities in a CTF environment, a process celebrated as a “\textit{very rewarding experience}” \cite{McDaniel2016}. Furthermore, the inherent difficulty of many challenges is instrumental in building resilience and persistence. By confronting and overcoming frustration in a low-stakes setting, students learn to treat setbacks as learning opportunities, developing the key attribute of “grit” that is strongly associated with long-term success \cite{Gough2024}.

In addition to individual cognitive growth, CTF environments are highly effective at developing crucial collaborative skills that mirror the modern workplace. While many competitions are structured for individuals, studies observe that students often form ad-hoc teams, with participants collaborating and helping each other when stuck \cite{Saddiqa2024}. Many educational programs intentionally leverage this by organizing team-based CTFs, which necessitate communication, delegation, and collaboration under pressure – consistently cited as major learning outcomes \cite{Li,VidenovikFili2024,Cherinka2018,Costa2023}. This peer interaction is a powerful learning mechanism; students explaining concepts to one another often internalize the material more deeply. This dynamic aligns with cognitive apprenticeship models, where more experienced students may take on informal mentorship roles, a process that benefits both mentor and mentee and enhances overall learning outcomes \cite{VidenovikFili2024,VidenovikVold2024}.

\subsection{Barriers to Effective and Equitable Implementation}
\label{sec4.4}

Despite these benefits, the literature also highlights numerous \textbf{barriers to implementing CTF competitions effectively in K-12 education}, particularly barriers that could hinder equitable participation. These challenges must be understood to inform a robust framework like ECAC. The identified barriers span practical resource constraints, pedagogical challenges, and issues of inclusivity. Table \ref{tab1} summarizes common barriers noted across studies, along with their impacts on CTF-based program implementation in schools. These challenges – ranging from lack of resources and teacher expertise to student skill gaps, design pitfalls, and inclusivity issues – must be addressed for successful integration of CTF-based learning.

\begin{longtable}{@{} p{0.2\textwidth} p{0.55\textwidth} p{0.2\textwidth} @{}}

\caption{Common Barriers to K-12 CTF Implementation.}
\label{tab1} \\

\toprule
\multicolumn{1}{c}{\textbf{Barrier Category}} &
\multicolumn{1}{c}{\textbf{Description and Impact}} &
\multicolumn{1}{c}{\textbf{Study}} \\
\midrule
\endfirsthead

\multicolumn{3}{c}{\tablename\ \thetable{} -- Continued from previous page} \\
\toprule
\multicolumn{1}{c}{\textbf{Barrier Category}} &
\multicolumn{1}{c}{\textbf{Description and Impact}} &
\multicolumn{1}{c}{\textbf{Study}} \\
\midrule
\endhead

\midrule
\multicolumn{3}{r@{}}{\textit{Continued on next page}} \\
\endfoot

\bottomrule
\endlastfoot

Infrastructure \& Resources &
High costs and technical requirements for running CTF environments. For example, operating a full cyber range (with virtual machines, networking, etc.) is ``\textit{expensive and requires technical staff in addition to extensive infrastructure},'' making it impossible for some schools to participate without external support. Many K-12 schools lack dedicated cybersecurity labs or sufficient hardware, creating a resource barrier. &
\cite{Ford2017,Blai2024,VidenovikFili2024,Namukasa2024,Manganello2024,Sreehari2023} \\
\addlinespace 

\textbf{Teacher Training \& Support} &
Shortage of educators with cybersecurity expertise. Many primary and secondary schools do not have teachers trained in cybersecurity, and teachers may lack confidence or materials to run CTF activities. This expertise gap means even willing schools struggle to implement competitions without significant professional development or external mentors. Ongoing support is needed so that instructors can guide students through challenges and address issues as they arise. &
\cite{Ford2017,Casey2023,VidenovikFili2024,McDaniel2016,Lazarov2025,Saddiqa2024,Manganello2024} \\
\addlinespace

\textbf{Curricular Constraints} &
Difficulty integrating CTFs into the standard curriculum. Cybersecurity is often not part of K-12 curricula, and adding a CTF module can face time and testing constraints. As a result, CTF activities are frequently conducted as extracurricular camps or clubs rather than during class. This limits access to those students who opt in or have after-school time, potentially exacerbating inequities (students without transportation or those unaware of the opportunity may be left out). &
\cite{Jin2018, Casey2023,Li, Ibrahim2020,Purbo2024,Rathore, Kikkas2020, MelloStark2020} \\
\addlinespace

\textbf{Student Preparedness Gaps} &
Wide variance in student baseline skills and knowledge. In any K-12 group, some students (especially younger ones or those from under-resourced backgrounds) may lack fundamental computing skills--e.g. understanding command-line interfaces or even managing files. Such gaps can make certain CTF challenges inaccessible or frustrating. Additionally, prior exposure to puzzle-solving or STEM enrichment varies, so a one-size-fits-all competition can overshoot the abilities of less-prepared students. Without adjustments or introductory lessons, these students may disengage early. &
\cite{Blai2024,Ibrahim2020,Rathore,Namukasa2024,Lazarov2025,Saddiqa2024,Manganello2024} \\
\addlinespace

\textbf{Competitive Format Challenges} &
The competition format itself can pose challenges. Time pressure and scoring create a high-pressure environment that not all students thrive in. Studies have noted that while competition\textit{ motivates some, it induces stress in others}. Students who are timid or fear failure might hold back or avoid participating. Similarly, common CTF design choices like penalizing the use of hints can discourage students from seeking help when they are stuck. This is counterproductive for learning and can turn the intended scaffolding (hints) into a source of anxiety. Finding the right balance in game mechanics is a notable implementation hurdle. &
\cite{Costa2023, Ford2017, Gough2024, McDaniel2016, Namukasa2024} \\
\addlinespace

\textbf{Ethical \& Real-World Concerns} &
Educators have raised concerns about the ethics and realism of CTF challenges. Some teachers worry that teaching offensive hacking techniques without proper context could lead to misuse or misunderstandings. There's a ``\textit{tainted image of hacking}'' in media, and without ethical guidance, students might not grasp the distinction between legal, ethical cybersecurity work and malicious hacking. Additionally, if challenges are too contrived or unrealistic, teachers fear students won’t see the real-world relevance (or worse, develop misconceptions about security). These concerns highlight the need to accompany CTF exercises with discussions of ethics and real-world applications--which, if not done, is a barrier for teacher buy-in and safe implementation. &
\cite{Casey2023,McDaniel2016,Kikkas2020,Saddiqa2024,Costa2023} \\
\addlinespace

\textbf{Diversity \& Inclusion} &
Ensuring equitable participation in CTFs is an ongoing challenge. As noted, cybersecurity clubs and competitions have historically attracted a narrow slice of students (often those who are already technologically inclined, and disproportionately male). Research indicates that engagement levels can differ by gender--e.g. one study found males enjoying the CTF experience more than females on average--suggesting that standard CTF formats may not appeal equally to all students. Other factors like socioeconomic gaps (access to computers at home, etc.) and school location (urban vs. rural opportunities) can also affect who participates. Without deliberate measures to broaden appeal and accessibility, CTF programs might leave out students from underrepresented groups, thus undermining the goal of inclusive cybersecurity education. &
\cite{Jin2018,Qusa2021,Casey2023,Gough2024,Purbo2024,VidenovikFili2024,Toccafondi2024,Namukasa2024,Costa2023,MelloStark2020} \\

\end{longtable}

In reviewing these barriers, it becomes clear that simply transplanting a collegiate or industry-style CTF into a K-12 setting is often ineffective or inequitable. Schools may lack the infrastructure, and students and teachers may lack the preparatory background, to fully benefit from a traditional CTF. The competitive format, if not adapted, can alienate a portion of the learners. Crucially, issues of equity mean that extra effort is needed to engage girls, underrepresented minorities, and less tech-confident students who might not initially see cybersecurity as “for them.” These findings underscore the importance of careful framework design (as we propose with ECAC) to mitigate barriers – for instance, by providing teacher training, using cost-effective platforms, scaffolding the learning experience, and embedding ethical guidance. The next theme directly addresses how modified CTF models compare to traditional ones in light of these needs.

\subsection{Comparative Insights: Traditional vs. Scaffolded/Inclusive CTF Models}
\label{sec4.5}

An overarching insight from the literature is the contrast between traditional CTF competitions and more scaffolded, inclusive CTF models designed for education. Traditional CTFs – such as those in cybersecurity tournaments or university hacking clubs – tend to be highly competitive, fast-paced, and \textbf{geared toward individuals with existing interest or expertise}. In a K-12 context, however, many researchers argue for a reimagined CTF format that aligns with pedagogical best practices and the diverse needs of younger learners \cite{Lazarov2025}. The ECAC framework builds on this concept of a scaffolded CTF. Key differences that emerge in the literature are summarized in Table \ref{tab2}. The latter introduces tailored support, broader content, and pedagogical structuring to transform the CTF from a purely competitive game into a learning-centric experience.

\begin{longtable}{@{} p{0.15\linewidth} p{0.25\linewidth} p{0.08\linewidth} p{0.25\linewidth} p{0.08\linewidth} @{}}

\caption{Comparison of traditional CTF competition vs. scaffolded/inclusive CTF models for K-12.}
\label{tab2} \\

\toprule
\multicolumn{1}{c}{\textbf{Aspect}} &
\multicolumn{1}{c}{\textbf{\begin{tabular}[c]{@{}c@{}}Traditional CTF Model \\ (Competition-Centric)\end{tabular}}} &
\multicolumn{1}{c}{\textbf{Study}} &
\multicolumn{1}{c}{\textbf{\begin{tabular}[c]{@{}c@{}}Scaffolded/Inclusive CTF \\ Model (Educational-Centric)\end{tabular}}} &
\multicolumn{1}{c}{\textbf{Study}} \\
\midrule
\endfirsthead

\multicolumn{5}{l}{\tablename\ \thetable{} -- Continued from previous page} \\
\toprule
\multicolumn{1}{c}{\textbf{Aspect}} &
\multicolumn{1}{c}{\textbf{\begin{tabular}[c]{@{}c@{}}Traditional CTF Model \\ (Competition-Centric)\end{tabular}}} &
\multicolumn{1}{c}{\textbf{Study}} &
\multicolumn{1}{c}{\textbf{\begin{tabular}[c]{@{}c@{}}Scaffolded/Inclusive CTF \\ Model (Educational-Centric)\end{tabular}}} &
\multicolumn{1}{c}{\textbf{Study}} \\
\midrule
\endhead

\midrule
\multicolumn{5}{r@{}}{\textit{Continued on next page}} \\
\endfoot

\bottomrule
\endlastfoot

\textbf{Intended Audience \& Prerequisites} &
Designed for self-selected enthusiasts; \textit{assumes prior knowledge} and high technical interest among participants. Often attracts students who already have some coding or cybersecurity experience. &
\cite{Ford2017,Gough2024,Purbo2024,Namukasa2024,Kikkas2020,Manganello2024} &
Designed for a \textit{broad range of skill levels}, including complete novices. Does not assume background knowledge--instead provides on-ramps for beginners (e.g. pre-CTF tutorials or explanations) to accommodate diverse participants. &
\cite{Blai2024, Casey2023, Ibrahim2020, McDaniel2016,Lazarov2025} \\
\addlinespace

\textbf{Challenge Difficulty Progression }&
\textit{Steep or variable difficulty}. Challenges may be presented in no particular order, sometimes leading beginners to hit a very hard problem first. Little built-in guidance on progression--participants are expected to self-select what to attempt. This can result in early frustration or disengagement for less experienced students. &
\cite{Ibrahim2020,Purbo2024,McDaniel2016, Lazarov2025} &
\textit{Gradually increasing difficulty}. Tasks are sequenced from basic to advanced to scaffold learning. Educators often include simple, “one-step” challenges initially to build confidence, then ramp up complexity as skills improve. This prevents early overwhelm and keeps students motivated as they progress. &
\cite{Casey2023,McDaniel2016, Lazarov2025,Costa2023,Saddiqa2024} \\
\addlinespace

\textbf{Hints and Support} &
\textit{Minimal support by design}. Traditional competitions often limit hints or attach significant point penalties to them, to emphasize problem-solving independence and competition fairness. Participants are largely on their own, and asking for help might even be against the rules. Instructor/mentor intervention is rare during play. &
\cite{Purbo2024,Lazarov2025,Costa2023} &
\textit{Embedded scaffolding and help}. Hints, tips, and even walkthrough snippets are integrated to assist learners at point-of-need, with little or no stigma for use. Teachers or mentors are actively involved, circulating to coach students who are stuck. The philosophy is that hints accelerate learning rather than spoil the game. Any hint penalties are tuned gently so as not to deter students from seeking help. &
\cite{Casey2023, Gough2024, Namukasa2024,McDaniel2016,Lazarov2025,Saddiqa2024} \\
\addlinespace

\textbf{Content Scope \& Context} &
\textit{Narrow, technical focus}. Challenges typically concentrate on specific technical domains (e.g. binary exploitation, reverse engineering, cryptography) and are often abstract puzzles not obviously tied to everyday contexts. Ethical or social implications are seldom addressed explicitly during the competition. &
\cite{Jin2018,Ibrahim2020,Gough2024,Purbo2024,Namukasa2024, Kuchar2023,Kikkas2020} &
\textit{Broader, contextualized content}. Educational CTFs often incorporate scenarios drawn from real life (e.g. social media privacy settings, website security in a school context) to make activities relevant to students’ lives. Topics may include not just hacking techniques but also concepts of digital ethics, online safety, and defensive skills, aligning with curricula and fostering holistic understanding. This broader approach has been found to appeal to a more general student audience, including those without a prior technical focus. &
\cite{Casey2023,Lazarov2025,Saddiqa2024,Costa2023} \\
\addlinespace

\textbf{Competitive vs. Collaborative Tone} &
Emphasis on \textit{competition}. Rewards speed and solo problem-solving; participants or teams compete to top the scoreboard. Traditional CTF culture valorizes being the first to solve and may discourage collaboration (except within a formal team in team events). Less-skilled participants can feel left behind in a fierce competition. &
\cite{Gough2024,Purbo2024,Cherinka2018,Namukasa2024,Lazarov2025} &
Emphasis on \textit{collaboration and learning}. While a competitive element exists, the inclusive model encourages teamwork and peer learning. Students are often observed helping each other and even solving challenges in groups. Many implementations incorporate team-based challenges or at least allow knowledge-sharing, reflecting a belief that collaboration can enhance learning outcomes. The competitive aspect is downplayed in favor of personal achievement and collective problem-solving. &
\cite{Casey2023, Gough2024,VidenovikFili2024, Namukasa2024, Saddiqa2024} \\
\addlinespace

\textbf{Duration \& Format} &
Often a \textit{single, time-bound event} (e.g. a 4-hour or one-day competition). Intense burst of activity with limited time for reflection or instructional interludes. Little continuity once the event ends (unless participants independently continue practice). &
\cite{Blai2024, Ibrahim2020, Purbo2024,McDaniel2016} &
Frequently a \textit{longer or repeated format}. Some programs integrate CTF challenges into a course or a multi-day workshop. For example, challenges might be spread over a week or semester, with periodic sessions that allow students to digest and revisit concepts. This format enables reflection, follow-up discussions, and iterative improvement. It transforms CTF from a one-off contest into a continuous learning tool. &
\cite{Qusa2021,Casey2023, Ibrahim2020, Costa2023,MelloStark2020} \\
\addlinespace

\textbf{Assessment of Outcomes} &
Success is measured by \textit{rank and flags captured}. The focus is on problem completion under pressure; learning is implicit but not formally assessed during the event. Typically, there’s no direct evaluation of what concepts students learned or how their skills improved – performance is gauged competitively. &
\cite{Purbo2024,Cherinka2018, McDaniel2016,Costa2023} &
Success is measured by \textit{learning gains and skill development}. Educators assess outcomes via pre/post quizzes, surveys, and practical skill demonstrations. For example, students’ improvement in knowledge (e.g. understanding of cybersecurity concepts or awareness of risks) is tracked; one program reported over 90\% of participants had increased understanding of key topics after the CTF. The competition is viewed as a means to an educational end, rather than the end itself. &
\cite{Ford2017,Blai2024,Qusa2021,Casey2023, Ibrahim2020, Toccafondi2024} \\

\end{longtable}

Across these dimensions, the \textbf{inclusive CTF model aligns more closely with educational best practices} and the cognitive apprenticeship approach – it meets students at their level, then guides them toward higher expertise with support and context at each step. For example, the importance of real-world context is repeatedly stressed: one study found that \textit{non-technical high school students engaged more and found tasks easier when challenges related to everyday life topics like social media privacy}, as opposed to abstract “hacking” tasks \cite{Saddiqa2024}. Teachers in that study emphasized linking concepts to the real world to sustain interest, and indeed participants came away with a heightened awareness of how cybersecurity impacts them (e.g. understanding how social media data can be misused) \cite{Saddiqa2024}. Such outcomes are less likely in a traditional CTF that lacks explicit teaching moments or relevant framing.

Likewise, \textbf{scaffolding measures} are a defining feature of the inclusive model. Educators frequently recommend and implement features like step-by-step tutorials, “\textit{hint buttons}”, and tiered challenge levels to accommodate the wide variance in student backgrounds \cite{McDaniel2016,Saddiqa2024}. In practice, these adjustments have proven necessary: in one deployment, many students struggled with unfamiliar terminology and tools, prompting calls for more in-platform guidance and even video walk-throughs for the uninitiated. Traditional CTFs rarely provide such aids – participants are expected to independently research and figure things out. But in a K-12 setting, without scaffolding, those with less experience often get stuck and disengage \cite{McDaniel2016}. Thus, the inclusive model’s philosophy is to \textbf{lower the floor without lowering the ceiling}: make it easier for beginners to get started (through hints, easier initial tasks, and coaching), while still offering harder problems that advanced students can stretch on. This approach keeps more students in the game and learning actively, addressing the equity gaps noted earlier. Notably, a scaffolded model also tends to incorporate explicit ethical guidance and a “white-hat” hacker mindset, which traditional CTFs take for granted. Studies point out that young participants need clear instruction in cybersecurity ethics – emphasizing defense, responsible disclosure, and the positive applications of hacking skills – to counter popular misconceptions \cite{Kikkas2020}. An inclusive CTF might include scenarios where students play the role of an investigator or defender (rather than a stereotypical outlaw hacker), thus reinforcing an ethical outlook while they solve challenges \cite{Casey2023,Ibrahim2020}. Traditional CTFs, in contrast, often focus solely on attack techniques and assume the participants bring their own ethical framework \cite{Kikkas2020}.

In sum, \textbf{the literature advocates for moving “beyond the flag” – adapting the CTF format to be more educationally inclusive}. Traditional CTF competitions are excellent for engaging motivated learners and identifying elite talent, but they can fall short as a teaching tool for broad audiences. The ECAC framework is informed by these comparative insights: it seeks to retain the engagement and authentic problem-solving of CTFs, while infusing scaffolding, contextualization, mentorship, and ethical training to support cognitive apprenticeship and inclusive skill development. By doing so, CTFs in K-12 can evolve from niche contests into powerful, inclusive learning experiences that build not only technical prowess but also the ethical and higher-order skills needed for the next generation of cybersecurity professionals \cite{Casey2023,McDaniel2016,Costa2023}. The findings from the literature make clear that such an evolution is both necessary and achievable, given thoughtful integration of the themes outlined above.

\section{Ethical-Cognitive Apprenticeship in Cybersecurity (ECAC) Framework}
\label{sec5}

\subsection{Integrating Cognitive Apprenticeship, Ethics, and Inclusion}
\label{sec5.1}

The ECAC framework is grounded in a synthesis of cognitive apprenticeship theory, ethical reasoning pedagogy, and inclusive educational principles. \textbf{Cognitive Apprenticeship (CA) }provides the backbone for skill acquisition and deep learning: it emphasizes making expert thinking visible and guiding novices from observation to independent mastery through methods like modeling, coaching, and scaffolding. This approach is especially suited to cybersecurity, where complex problem-solving benefits from expert demonstration and guided practice in authentic contexts. \textbf{Ethical Reasoning }pedagogy adds a crucial moral dimension. Given the dual-use nature of cybersecurity skills – which can be applied for defense or offense – an explicit focus on ethics is indispensable. Integrating theories of moral development and ethical decision-making into the learning design ensures that students grapple with questions of right and wrong as they acquire technical skills, cultivating “conscious and responsible digital citizens” rather than just capable hackers. \textbf{Inclusive Pedagogy} underpins ECAC’s commitment to broadening participation. Traditional cybersecurity competitions often appeal to a narrow subset of students and can alienate novices or those from underrepresented groups. ECAC draws on inclusive education strategies – such as culturally relevant examples, collaborative learning, and differentiated scaffolding – to engage diverse learners and ensure no prior experience or demographic background is a barrier to entry. By intertwining these foundations, ECAC harnesses the motivational power of gamified competition while mitigating its known shortcomings through structured learning and ethical context. This theoretically coherent blend provides a robust basis for a framework that aims to produce not only skilled problem-solvers, but ethical and empowered participants in the digital world.

\subsection{Core Design Principles of the ECAC Framework}
\label{sec5.2}
Building on these foundations, the ECAC framework is guided by several core design principles (Fig. \ref{fig1}) that shape its curriculum and implementation:

\begin{figure}[!h]
\centering{}
\includegraphics[width=0.8\textwidth]{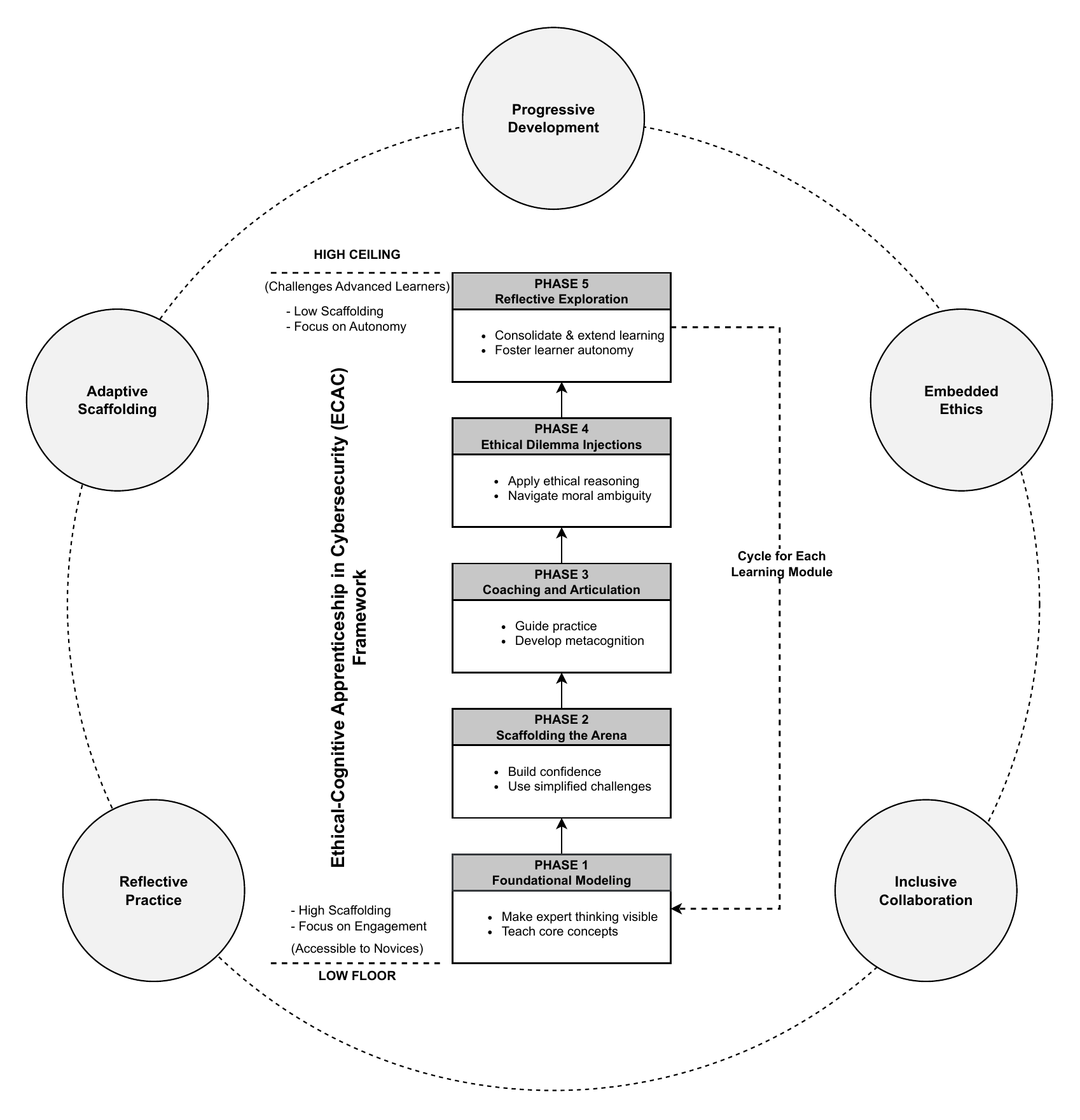}
\caption{ECAC Framework.\label{fig1}}
\end{figure}  

\begin{itemize}
\item \textbf{Progressive Development:} Skills and challenges develop in a carefully sequenced manner, matching learners’ cognitive stages. Cybersecurity concepts and tasks are introduced in increasing complexity as students mature, preventing cognitive overload for younger learners and ensuring sustained challenges for older students. This developmental progression ensures each new challenge builds on prior knowledge, fostering steady growth rather than frustration or boredom.
\item \textbf{Embedded Ethics:} Ethical considerations are woven directly into activities and challenges, rather than treated as an add-on topic. Many “flags” in ECAC-designed competitions require ethical judgments or reflections in addition to technical solutions, prompting students to confront moral dilemmas as part of the problem-solving process. By normalizing ethical inquiry within technical tasks, the framework reinforces that knowing \textit{how} to do something in cybersecurity must be accompanied by understanding \textit{should} it be done.
\item \textbf{Inclusive Collaboration:} ECAC emphasizes teamwork and peer learning to make cybersecurity learning more inclusive and community-oriented. Challenges are often tackled in groups, and activities are structured to value diverse contributions (e.g. brainstorming, strategy, technical execution) so that students with different skill levels and backgrounds can all participate meaningfully. A collaborative environment not only mirrors real-world cybersecurity practices, but also helps demystify content for novices and supports learners who might feel isolated in competitive, high-pressure settings.
\item \textbf{Reflective Practice:} Beyond solving challenges, ECAC embeds regular reflection and articulation of learning. After each competition activity or module, students engage in debriefs, write-ups, or discussions to articulate what strategies they used, examine mistakes, and compare approaches with peers or experts. This reflective practice cements learning, develops metacognitive awareness, and reinforces ethical thinking by having students consider the implications of their actions and decisions in hindsight.
\item \textbf{Adaptive Scaffolding:} The framework employs dynamic scaffolding – temporary supports, hints, and structured guidance – that adapts to learners’ needs and gradually fades as competence grows. Early on, students might receive step-by-step hints, simplified “unplugged” exercises, or pre-configured safe environments to practice in. As they become more proficient, these supports are removed or made optional, challenging students to take increasing ownership of problem-solving. This adaptive scaffolding ensures that every student, from beginner to advanced, can engage productively with cybersecurity challenges within their zone of proximal development.
\end{itemize}
These principles collectively ensure that ECAC’s implementation remains \textbf{developmentally appropriate, ethically conscientious, collaborative, reflective, and responsive} to individual learning needs. They set ECAC apart from one-size-fits-all or purely competition-driven approaches by ensuring a balanced, education-first design of cybersecurity competitions.

\subsection{Phase-by-Phase Framework Structure}
\label{sec5.3}
ECAC operationalizes the above principles through a series of phases that align with the cognitive apprenticeship model and embed ethical inquiry at each step. The phases represent a cycle of learning within a cybersecurity competition module or curriculum unit, progressing from introduction to independent practice, all infused with ethical context. The five key phases (Fig. \ref{fig1}) of the ECAC framework are: \textbf{Foundational Modeling, Scaffolding the Arena, Coaching and Articulation, Ethical Dilemma Injections,} and \textbf{Reflective Exploration}. Each phase is described below.

\subsubsection{Foundational Modeling}
\label{sec5.3.1}
The Foundational Modeling phase establishes the necessary groundwork for learners by providing explicit instruction on core cybersecurity concepts before any competitive engagement. In this initial stage, expert instructors make their tacit knowledge explicit for novices by demonstrating technical procedures and verbalizing the thought processes behind them. This modeling makes the cognitive strategies typically hidden in cybersecurity work visible. For example, a teacher might walk through a simple capture-the-flag puzzle, narrating their steps to show how to systematically break down the problem and select the right tools.

Crucially, this phase integrates ethical reasoning directly with technical competence. As instructors solve problems, they also model ethical decision-making, perhaps by discussing professional codes of conduct or voicing considerations like, “\textit{I won’t try to break the rules to get the flag, because…}”. This approach can be adapted for different age groups, using everything from interactive stories for younger students to live demos of network tools for older ones. To foster an inclusive environment, this phase also benefits from using diverse instructors who acknowledge that different backgrounds and perspectives contribute to effective problem-solving. The ultimate goal is to provide students with a concrete cognitive and ethical model that serves as a reference point for their own practice in subsequent phases.

\subsubsection{Scaffolding the Arena}
\label{sec5.3.2}
The Scaffolding the Arena phase transforms the traditional, often "sink-or-swim" CTF competition into a structured learning environment designed for progressive skill development. The goal is to systematically build student competencies by creating scaffolded experiences rather than presenting a series of isolated problems. The "arena," which could be an online platform or a classroom activity, is intentionally prepared with supportive features to help novices engage with complex tasks step-by-step. For instance, challenges are designed with a "low floor, high ceiling" approach, starting with accessible entry-level tasks to guarantee an early "win" and build confidence before moving to more advanced problems. Unlike typical competitions where hints are minimal or penalized, this model provides just-in-time learning resources, contextual hints, and pre-configured tools without penalty, allowing students to focus on concepts rather than setup obstacles.

This phase also redesigns the social and ethical context of the competition. Collaboration is encouraged through team-based challenges and peer mentoring opportunities, and assessment is modified to reward the learning process itself, such as documenting problem-solving approaches or helping teammates. Ethical reflection is woven directly into the challenges, with prompts that ask students to consider the implications of their actions, such as, "\textit{what is the right thing to do with this information?}". This scaffolding is not permanent; it is gradually removed as learners' competencies increase, making tasks more open-ended and requiring more independence. By carefully engineering the arena in this way, students can build skills and confidence in a supportive environment before tackling fully autonomous challenges.

\subsubsection{Coaching and Articulation}
\label{sec5.3.3}
In the Coaching and Articulation phase, learning is solidified through guided practice and metacognitive reflection. While students work through challenges, instructors transition to the role of a coach, circulating to monitor progress and provide real-time, tailored feedback. This coaching is not about giving answers, but rather about preventing unproductive struggle by asking probing questions, giving strategic hints, or suggesting alternative approaches. In tandem with receiving guidance, students are required to articulate their problem-solving processes. This makes their thinking visible and helps them develop crucial metacognitive skills.

For example, a coach might prompt a team to explain their technical reasoning by asking, “\textit{Can you walk me through how you’re trying to decode that cipher and why?}” This process is also applied to ethical dilemmas, with questions like, “\textit{Is there an easier way to get that information, and is it okay to do so?}” to spur discussion. Students may be asked to document their approaches or participate in peer discussions, and in some cases, advanced students can even serve as peer coaches. By requiring students to verbalize their technical and ethical reasoning, this phase reinforces understanding, reveals misconceptions, and prepares them for more complex, independent work.

\subsubsection{Ethical Dilemma Injections}
\label{sec5.3.4}
A distinguishing feature of ECAC is the deliberate insertion of ethical dilemmas into the flow of challenges. In this phase, as students become comfortable with the technical routine of a competition, the framework “injects” a scenario twist or decision point that foregrounds an ethical question. This could be a challenge where the straightforward way to capture a flag involves an action that is legally or morally questionable, forcing students to pause and consider alternatives. For example, partway through a network security CTF scenario, students might discover they could obtain a flag by exploiting a vulnerability that would violate someone’s privacy; the dilemma is whether to use that exploit or find a more ethical solution. Such injections are often implemented via story elements or explicit tasks – e.g. “\textit{Option A: Access the admin account using the leaked password you found (Flag if you do so). Option B: Do not use the password and instead report the issue – write what you would do next.}” The goal is to have students confront the consequences of “bending the rules” within a safe learning context. Instructors facilitate discussions around these moments, guiding students to articulate why a particular action is or isn’t acceptable. This mirrors real-world cybersecurity ethics (such as responsible disclosure vs. unauthorized access) and ensures that ethical reasoning is not an afterthought but a core challenge to be solved. By navigating these dilemmas, students practice moral decision-making alongside technical problem-solving, learning to prioritize ethical principles even under competitive pressure.

\subsubsection{Reflective Exploration}
\label{sec5.3.5}
The final phase of the ECAC cycle centers on reflection and extension of learning. After the main challenges (including any ethical dilemma components) are completed, learners engage in structured reflection. They review their performance, discuss what strategies worked or failed, and critically examine their decisions – both technical and ethical. This might involve writing a short “\textit{after-action report}” on how they solved a particular challenge and how they dealt with the ethical issues, or a guided class discussion comparing different teams’ approaches and moral reasoning. Reflection solidifies the lessons learned and connects the competition experience to broader concepts and real-life practices. Following reflection, the framework encourages exploration beyond the set challenges. Students are prompted to take their new skills and curiosity further: for example, they might be asked to devise a new challenge or experiment with a tool not used during the competition, or to research a recent cybersecurity incident related to the competition theme. In this way, learning becomes self-driven – an important element of cognitive apprenticeship’s emphasis on learner autonomy in later stages. Importantly, this exploration is still tied back to ethics and responsibility. Students could be tasked with exploring how a vulnerability they exploited in a challenge is handled in industry, or how misinformation (if that was the theme) affects their own community, and then reflect on the ethical dimensions of those real-world connections. The Reflective Exploration phase thus consolidates knowledge, promotes long-term retention, and encourages learners to continue investigating cybersecurity topics with a critical, ethical mindset even after the formal activity ends.

\subsection{ECAC Framework Summary}
\label{sec5.4}
In summary, the ECAC framework marries the staged guidance of cognitive apprenticeship with the engaging format of cybersecurity competitions, all within a strong ethical and inclusive envelope.
To provide a concise view, Table \ref{tab3} compares the ECAC framework with traditional K-12 cybersecurity learning approaches. This highlights ECAC’s unique features, such as its continuous mentoring, embedded ethics, and inclusivity, versus the more ad-hoc and purely technical focus of typical capture-the-flag events.

\begin{longtable}{@{} p{0.2\linewidth} p{0.38\linewidth} p{0.38\linewidth} @{}}

\caption{Comparison of a typical CTF competition model with the ECAC framework approach in K-12 cybersecurity education.}
\label{tab3} \\

\toprule
\multicolumn{1}{c}{\textbf{Aspect}} &
\multicolumn{1}{c}{\textbf{Traditional CTF Model}} &
\multicolumn{1}{c}{\textbf{ECAC Framework}} \\
\midrule
\endfirsthead

\multicolumn{3}{l}{\tablename\ \thetable{} -- Continued from previous page} \\
\toprule
\multicolumn{1}{c}{\textbf{Aspect}} &
\multicolumn{1}{c}{\textbf{Traditional CTF Model}} &
\multicolumn{1}{c}{\textbf{ECAC Framework}} \\
\midrule
\endhead

\midrule
\multicolumn{3}{r@{}}{\textit{Continued on next page}} \\
\endfoot

\bottomrule
\endlastfoot

\textbf{Learning Structure} &
One-off competitions, minimal scaffolding or progression; often outside formal curriculum. &
Integrated into curriculum as a longitudinal pathway with staged progression and continuous skill building. \\
\addlinespace

\textbf{Focus of Challenges} &
Primarily technical “flag capturing,” emphasis on point-scoring and competition. &
Holistic tasks combining technical problem-solving with ethical decision points and real-world context. \\
\addlinespace

\textbf{Pedagogical Support }&
Learn by doing with little guidance; experts may design challenges but do not actively coach during competition. &
Cognitive apprenticeship methods applied: teachers/mentors model solutions, coach during challenges, scaffold tasks, and prompt reflection. \\
\addlinespace

\textbf{Ethics Integration} &
Usually absent or confined to separate talks; winning is prioritized over moral considerations. &
Ethics is deeply interwoven: some flags require ethical reasoning, and students engage in guided ethical discussions as part of competition. \\
\addlinespace

\textbf{Inclusivity \& Access} &
Tends to attract students who are already confident or experienced; can be intimidating to novices and marginalized groups. &
Designed for broad inclusion: beginner-friendly entry points, team collaboration, culturally relevant scenarios, and adaptive difficulty to welcome diverse participants. \\
\addlinespace

\textbf{Outcomes Emphasized} &
Technical skills, quick problem-solving; winners gain recognition but learning depth varies. &
Deeper learning outcomes: sustained problem-solving ability, critical thinking, ethical decision-making skills, teamwork, and long-term interest in cybersecurity. \\

\end{longtable}

As shown above, ECAC’s design directly addresses many weaknesses of traditional models by providing a structured, supportive, and ethics-focused learning environment. In the next section, we illustrate how ECAC works in practice through scenarios and examine its effectiveness for different types of learners and contexts.

\section{ECAC in Action: A Staged K-12 Implementation}
\label{sec6}
\begin{figure}[!h]
\centering{}
\includegraphics[width=\textwidth]{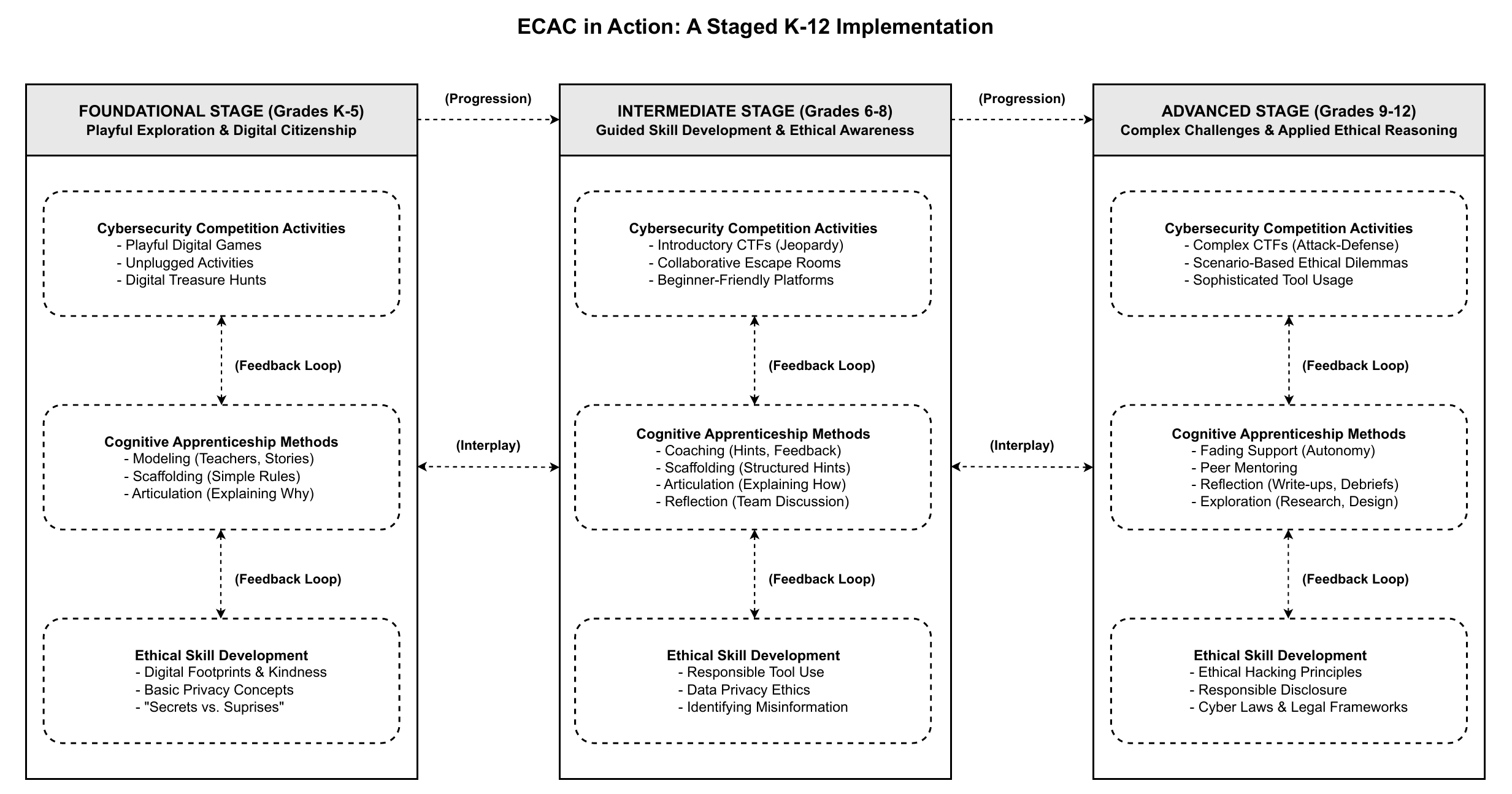}
\caption{ECAC in Action: A Staged K-12 Implementation.\label{fig2}}
\end{figure}

To translate the ECAC framework from a theoretical model into a practical pedagogical guide, this section illustrates how its five phases can be operationalized across the K-12 educational spectrum (Fig. \ref{fig2}). The implementation is staged to be developmentally appropriate, ensuring that concepts, challenges, and ethical inquiries align with the cognitive and emotional maturity of learners at each level. The complexity of activities and the application of the ECAC phases evolve from playful exploration in elementary grades to sophisticated, autonomous problem-solving in high school.

\subsection{Foundational Stage (Grades K-5): Playful Exploration and Digital Citizenship}
\label{sec6.1}

The primary objective at this stage is to introduce fundamental concepts of online safety and responsible digital behavior through engaging, non-intimidating activities. The focus is less on technical skill and more on building a strong foundation of digital citizenship through guided play.

\textbf{Illustrative Activities:} Learning occurs through story-driven digital games and "unplugged" activities that teach cybersecurity concepts through narrative and play. Examples include interactive games that model safe website identification or offline role-playing scenarios about creating "secret codes" to introduce basic cryptographic thinking, adapting concepts from initiatives for a younger audience.

\textbf{ECAC Phases in Practice:}
\begin{itemize}
\item \textit{Foundational Modeling:} This phase is paramount. Teachers and game narratives explicitly model safe online behaviors and simple problem-solving, such as what to do when encountering a stranger online. For instance, in a "Digital Detectives" game, a teacher might think aloud: "\textit{This website is asking for my real name. Good detectives know we shouldn't share personal information. I'll use my codename instead.}" This makes an expert's internal monologue explicit.
\item \textit{Scaffolding the Arena:} The arena is scaffolded by creating highly structured, low-risk learning environments. In practice, this means the game interface is simplified, offering only three clear, icon-based choices (e.g., \textit{"Share Info," "Ask a Trusted Adult," "Go Back"}), which prevents confusion and allows young learners to succeed.
\item \textit{Coaching and Articulation:} Coaching is facilitated through guided discussions, while articulation is encouraged by having students explain their reasoning in simple terms. For example, when a student hesitates, the teacher acts as a coach, asking a guiding question: "\textit{What do we know about strangers online? What did our detective rules say?}". Afterwards, the student is prompted to explain their choice to a partner, articulating the "\textit{why}" behind their action.
\item \textit{Ethical Dilemma Injection:} Ethical considerations are introduced as simple, narrative-based choices. A pop-up might appear in the game stating: "\textit{Another player offers you a cheat code to beat the game if you share your character's 'secret key'. What do you do?}" This presents a relevant ethical choice within a safe context.
\item \textit{Reflective Exploration:} Reflection connects in-game actions to broader principles. After the game, the teacher can lead a discussion: "\textit{Why is it a bad idea for everyone in the game if players start sharing their secret keys?}" This helps students explore and generalize learning from the activity.
\end{itemize}

\subsection{Intermediate Stage (Grades 6-8): Guided Skill Development and Ethical Awareness}
\label{sec6.2}
This stage builds upon the K-5 foundation by introducing more specific cybersecurity concepts and basic technical skills through guided, collaborative, and CTF-like activities, alongside a growing awareness of ethical responsibilities.

\textbf{Illustrative Activities: }This stage introduces introductory, Jeopardy-style CTFs using beginner-friendly platforms like CTFd (\url{https://ctfd.io} [Access date: June 15, 2025]). Challenge categories are age-appropriate, covering topics like simple ciphers, steganography, or basic Open Source Intelligence (OSINT). Collaborative activities like virtual "\textit{escape rooms}" that require solving a series of cyber-related puzzles are also highly effective.

\textbf{ECAC Phases in Practice:}
\begin{itemize}
\item \textit{Foundational Modeling:} Instructors continue to make their thinking visible while demonstrating more technical tasks. For example, when introducing cryptography, a teacher models the process for a Caesar cipher using a cipher wheel, narrating their steps: "\textit{I see the hint says 'Shift 3'. That means I'll line up 'A' with 'D' and then substitute each letter. Let's try the first word together.}"
\item \textit{Scaffolding the Arena:} The CTF "arena" is scaffolded with challenges that gradually increase in difficulty and feature integrated support systems. For instance, the CTF platform can include a prominent "Hint" button that provides progressively more specific clues without a major point penalty, encouraging help-seeking as a learning strategy rather than failure.
\item \textit{Coaching and Articulation:} Educators transition fully into the role of a coach, providing tailored guidance and prompting students to articulate their problem-solving processes. As a team gets stuck on a harder cipher, the coach might ask, "\textit{You've tried a Caesar cipher and it didn't work. What does the pattern of repeated letters tell you? Have you considered frequency analysis?}". The team is then asked to document their final solution, articulating their methods.
\item \textit{Ethical Dilemma Injection:} Ethical dilemmas are woven directly into technical challenges. For example, after students decrypt a message, they find it contains a fictional student's private, embarrassing diary entry. The flag is not the text itself but the answer to the question: "\textit{What is the ethical next step? A) Post it for laughs, B) Keep it to yourself, C) Report the exposed diary to a teacher.}" This forces a moral judgment to complete the task.
\item \textit{Reflective Exploration:} Reflection is formalized through structured debriefs where students compare problem-solving approaches and connect them to real-world contexts. The discussion moves beyond the technical solution to ethics: "\textit{Now that you know how to decrypt simple messages, how could this skill be used for good? How could it be misused? What are the real-world consequences of a data leak?}".
\end{itemize}

\subsection{Advanced Stage (Grades 9-12): Complex Challenges and Applied Ethical Reasoning}
\label{sec6.3}
The advanced stage focuses on developing sophisticated technical skills, critical thinking, and nuanced ethical reasoning through complex, scenario-based CTFs that prepare students for higher education and potential careers.

\textbf{Illustrative Activities:} Competitions become more complex, introducing topics like web exploitation (e.g., basic XSS, SQL injection), network forensics, and even simplified Attack-Defense scenarios within controlled cyber ranges or virtualized labs. Students use a wider array of standard cybersecurity tools, and activities include timely scenarios like securing Internet of Things devices.

\textbf{ECAC Phases in Practice:}
\begin{itemize}
\item \textit{Foundational Modeling:} Modeling is performed by experts on complex tasks, often demonstrating professional workflows. For example, an industry professional or teacher demonstrates how to use \textit{Wireshark} to isolate specific traffic and reassemble a file from network packets, explaining the "\textit{why}" behind each filter used.
\item \textit{Scaffolding the Arena (with Fading): }Scaffolding becomes dynamic and fades over time to promote autonomy. Instead of providing step-by-step instructions, a challenge description might only state, "\textit{The attacker exfiltrated a file using FTP. Analyze capture.pcap to find it.}" Students are expected to research the necessary commands and protocols on their own, encouraging independent learning.
\item \textit{Coaching:} The instructor's role shifts to that of a high-level coach or mentor, asking probing questions to guide critical thinking rather than procedural steps. A coach might ask a team, "\textit{Your SQL injection query failed. What do you know about how the database might be sanitizing inputs? What's another character you could use to escape the string?}".
\item \textit{Ethical Dilemma Injection: }Dilemmas become sophisticated, reflecting the gray areas of the profession. While analyzing network traffic, students might find not only the corporate data they were assigned to find but also the personal, non-work-related Browse history of an employee. A section of the required report would then ask them to describe how they would ethically handle this incidental finding and justify their reasoning based on professional codes of conduct.
\item \textit{Articulation and Reflective Exploration:} These phases become more formal and central to assessment. The culminating task is to submit a formal "\textit{Incident Report}" or "\textit{CTF Write-up}". This task serves as \textbf{articulation} (detailing their technical process) and \textbf{reflection} (discussing the breach's impact and the ethical choices they made). The \textbf{exploration} phase then prompts students to research a real-world data breach that used similar techniques and present on the legal and societal consequences, encouraging self-driven, long-term learning.
\end{itemize}

\section{Validation of the ECAC Framework}
\label{sec7}

Having outlined the ECAC framework's structure and rationale, this section serves to validate its conceptual soundness and potential utility through non-empirical methods common in educational research. This approach is essential for establishing a theoretical framework's coherence and feasibility prior to large-scale empirical testing. The validation proceeds through three distinct stages. First, it demonstrates how the framework's design directly addresses key implementation barriers identified in the literature. Second, it uses scenario-based analysis with student archetypes to illustrate the framework's pedagogical flexibility for diverse learners. Finally, it aligns the framework with the successful, real-world \textit{GenCyber} program, using it as a case study to validate ECAC's practical applicability.

\subsection{Addressing Key Barriers to Implementation}
\label{sec7.1}
A primary method for validating a new framework is to demonstrate its capacity to solve existing, well-documented problems. The ECAC framework was specifically designed to mitigate the persistent barriers that hinder the effective and equitable integration of cybersecurity competitions into K-12 settings. This section analyzes how each core component of the ECAC framework serves as a direct, evidence-based solution to these challenges.

\begin{itemize}
    \item \textbf{Barrier: }\textbf{The Teacher Expertise Gap.} The literature consistently identifies a critical gap between the specialized knowledge required to teach cybersecurity and the training of the current K-12 educator workforce. Many teachers feel ill-prepared and lack the confidence to lead complex technical activities, creating a significant bottleneck for program adoption.
\end{itemize}

\textbf{ECAC Solution:} The framework directly addresses this by reframing the educator's role from a "content expert" to a "Lead Learner" or expert facilitator. ECAC implementation does not require teachers to be cybersecurity gurus; instead, professional development focuses on pedagogical skills like guiding inquiry, facilitating collaboration, and mentoring. The framework provides structured, scaffolded curricula with pre-designed challenges and embedded supports (Phase 2: Scaffolding the Arena), which reduces the burden on teachers to create technical content from scratch.

\begin{itemize}
    \item \textbf{Barrier: Resource and Infrastructure Inequality.} Schools, particularly those in under-resourced or rural areas, often lack the funding, technology, and high-speed internet access required for many cybersecurity competitions. This resource inequality creates significant barriers to access and exacerbates the digital divide.
\end{itemize}

\textbf{ECAC Solution: }The framework is designed for \textbf{scalable and adaptive implementation }that directly addresses resource constraints through its phased structure. Phase 1: Foundational Modeling intentionally uses accessible, low-tech formats like interactive storytelling, role-playing, and "unplugged" paper-based activities that require minimal infrastructure. As students progress, Phase 2: Scaffolding the Arena is also designed for flexibility; the "arena" can be a simple classroom activity or a free, open-source online platform, avoiding the need for expensive, dedicated cyber ranges at introductory levels. The later phases focusing on coaching, ethical discussion, and reflection (Phases 3, 4, and 5) are primarily pedagogical and less dependent on high-cost technology. For more advanced stages, ECAC emphasizes the use of free, open-source, or cloud-based platforms (like \textit{CTFd}) to minimize costs. This staged, flexible approach allows schools to adopt and scale the program according to their specific resource availability, ensuring broader and more equitable access.

\begin{itemize}
    \item \textbf{Barrier: Equity, Inclusivity, and Student Engagement.} Without intentional design, CTFs can be exclusionary. The competitive, often narrowly-focused "hacker" culture can alienate girls and underrepresented students. Furthermore, a steep learning curve and lack of support for novices can lead to frustration and disengagement, reinforcing existing gaps between students with and without prior experience.
\end{itemize}

\textbf{ECAC Solution:} The principle of Inclusive Collaboration is woven throughout the framework's design. ECAC promotes team-based activities where diverse contributions – such as research, communication, and strategic thinking, not just technical execution – are valued. Phase 2: Scaffolding the Arena provides a "low floor, high ceiling" design, ensuring beginners can experience early success while advanced students remain challenged. This approach has been empirically shown to increase engagement among underrepresented groups. By emphasizing collaboration over pure competition and using narrative-driven challenges with culturally relevant scenarios, ECAC broadens the appeal of cybersecurity and creates a more welcoming environment for all learners.

\begin{itemize}
    \item \textbf{Barrier: Lack of Integrated Ethical Training.} A significant risk in cybersecurity education is teaching powerful technical skills without a strong ethical foundation. Many programs treat ethics as an afterthought or a separate lecture, disconnected from the hands-on practice, which can fail to instill a durable sense of professional responsibility.
\end{itemize}

\textbf{ECAC Solution:} The ECAC framework is unique in its deep, systematic integration of ethics. This is not an add-on but a core component, realized through Phase 4: Ethical Dilemma Injections. In this phase, ethical questions are embedded directly into the technical challenges, forcing students to make moral judgments to "capture the flag". This process is reinforced through Phase 3: Coaching and Phase 5: Reflective Exploration, where students must articulate and justify their ethical choices. By making ethics an inseparable part of the problem-solving process, ECAC cultivates a mindset where "\textit{how}" and "\textit{should}" are as important as "\textit{can}".

\subsection{Scenario-Based Validation with Student Archetypes}
\label{sec7.2}

To evaluate ECAC’s design, we can envision how it caters to the varied types of learners commonly found in K-12 settings. By considering three archetypal students – a novice facing barriers, a technically skilled but ethically challenged "rule-bender," and a collaborative learner who thrives in groups – we can demonstrate how the framework's integrated design supports a wide range of needs, transforming their educational experiences from potentially negative to overwhelmingly positive.

\textbf{Scenario 1: The Novice Student}
\begin{itemize}

\item \textbf{Profile:} A student with no prior exposure to cybersecurity, who is curious but easily discouraged by highly technical or competitive environments. This student may lack confidence in their computer skills and face access barriers, such as not having a dedicated computer or high-speed internet at home, making them hesitant to join a "tech club".
\item \textbf{Traditional CTF Experience:} In a typical, unmodified CTF event, this novice would likely feel out of their depth. The competitive atmosphere can be intimidating, and the challenges often assume background knowledge. Faced with jargon-laden problems and minimal guidance, the student would likely disengage, which can dampen their interest in the field. The need for out-of-class practice would be an insurmountable barrier, effectively excluding them.

\item \textbf{ECAC Framework Experience:} The novice's journey is fundamentally different under ECAC. 
\begin{itemize}

\item \textit{Phase 1 (Foundational Modeling)} begins with an instructor presenting a simple, relatable cyber scenario and modeling how to think about the problem, which demystifies the process. This phase intentionally uses "unplugged" activities to ensure no student is left behind due to a lack of resources.
\item In \textit{Phase 2 (Scaffolding the Arena)}, the student encounters a "low floor" environment with a series of gradually challenging mini-tasks and plentiful hints, allowing them to gain small wins that boost confidence.
\item During \textit{Phase 3 (Coaching \& Articulation)}, a teacher acts as a supportive coach, quietly assisting and prompting the student to articulate how they solved a puzzle, which reinforces their new knowledge.
\item This supportive journey, with its emphasis on in-class, collaborative work, keeps the novice engaged and growing, transforming a tentative beginner into a motivated learner.
\end{itemize}
\end{itemize}

\textbf{Scenario 2: The "Rule-Bender" Student}
\begin{itemize}
\item \textbf{Profile:} A tech-savvy student who loves to push boundaries and is more interested in "winning" or showing off skills than following rules. They might already know some hacking tricks and could be the type to attempt cheating if it provides an edge.
\item \textbf{Traditional CTF Experience:} This student might thrive technically in a standard CTF but receive little ethical guidance. The competitive setting might even tacitly reward them for finding an unintended shortcut. Their takeaway could be to "do whatever it takes to get the flag," reinforcing a cavalier attitude toward cybersecurity ethics.

\item \textbf{ECAC Framework Experience:} ECAC proactively addresses this behavior through its integrated ethical structure. 
\begin{itemize}
\item \textit{Phase 4 (Ethical Dilemma Injections) }is central, presenting challenges where the straightforward way to capture a flag is legally or morally questionable. For example, a scenario might allow a flag to be obtained via an unsafe method but then pose a follow-up question asking why that method was problematic, forcing a consideration of consequences.
\item During \textit{Phase 3 (Coaching \& Articulation)}, if the student attempts an approach that violates the intended norms, the coach intervenes. Rather than simply penalizing, the coach turns it into a learning moment, prompting a discussion about the rules and their real-world importance.
\item Through this continuous exposure to ethics-in-action, the student learns to channel their ingenuity within ethical constraints, such as reporting a vulnerability responsibly rather than exploiting it maliciously.
\end{itemize}
\end{itemize}

\textbf{Scenario 3: The Collaborative Learner}
\begin{itemize}
\item \textbf{Profile:} A student who learns best with others, enjoying discussion and joint problem-solving. They excel in communication but may feel alienated by solo competition, a trait often seen in students from groups underrepresented in tech who value a supportive team environment.
\item \textbf{Traditional CTF Experience:} Even in team-based CTFs, a few "experts" on the team often take over, relegating others to spectator roles. Without structured interaction, a social learner can feel underutilized, and research has noted that without facilitation, voices from marginalized groups may be drowned out.
\item \textbf{ECAC Framework Experience:} Collaboration is intentionally built into every layer of ECAC's design. 
\begin{itemize}
\item Challenges are designed for teamwork, and instructors are trained to facilitate equitable participation.
\item During \textit{Phase 3 (Coaching \& Articulation)}, teams are encouraged to assign roles, allowing the collaborative learner to act as a coordinator or explainer. This student thrives when they must articulate the team's progress or explain a solution, learning deeply by teaching others.
\item Group dialogues around \textit{Phase 4} ethical dilemmas become a moment for this student to shine, helping the team weigh different viewpoints.
\item By valuing communication as much as technical prowess, ECAC transforms the competition from a zero-sum race into a community learning endeavor, which research suggests can improve outcomes and make the field more welcoming to all students.
\end{itemize}
\end{itemize}

\subsection{Use Case Validation: The GenCyber Program as an ECAC Model}
\label{sec7.3}
A powerful method for validating a conceptual framework is to demonstrate its alignment with existing, successful real-world programs. The \textbf{GenCyber program} (\url{https://public.cyber.mil/gencyber} [Access date: June 15, 2025]), a national initiative jointly sponsored by the National Security Agency (NSA) and the National Science Foundation (NSF), provides an excellent use case. While GenCyber camps \cite{Ladabouche2016} do not explicitly use the ECAC name, their documented pedagogical structure and core principles strongly mirror the ECAC framework's five phases, validating its practical feasibility and effectiveness (\url{https://sites.google.com/mocs.utc.edu/gencyber/about} [Access date: June 15, 2025]).
\begin{itemize}

\item \textbf{Phase 1:} Foundational Modeling: GenCyber camps typically begin by introducing core concepts before any hands-on activities. Instructors model key ideas, such as the CIA Triad (Confidentiality, Integrity, Availability), and demonstrate safe online practices, providing students with a solid conceptual foundation before they engage with technical tools (\url{https://tinyurl.com/GenCyberCore} [Access date: June 15, 2025]).
\item \textbf{Phase 2: Scaffolding the Arena:} The curriculum in these camps is highly scaffolded to be accessible and engaging. Many camps use "unplugged" activities to introduce complex topics simply, followed by a progression to guided labs \cite{hernandez2020engaging}. This "low floor, high ceiling" approach ensures that students with no prior experience can participate successfully, a core tenet of ECAC's design that aligns with GenCyber's goal to "excite a new generation of learners” \cite{Ladabouche2016}.
\item \textbf{Phase 3: Coaching and Articulation:} GenCyber camps rely heavily on a coaching model, with university faculty and student mentors circulating to provide individualized guidance during labs. Furthermore, many camps culminate in student presentations or the development of lesson plans, requiring students to articulate what they have learned, which is a key component of the ECAC cycle.
\item \textbf{Phase 4: Ethical Dilemma Injection:} A core tenet of the program is its focus on ethics, aligning perfectly with the 'GenCyber First Principles' which mandate that ethics be an integral part of the curriculum. Camp curricula often include specific modules on correct and safe online behavior and the ethics of hacking. These discussions, integrated alongside technical labs, serve the same function as ECAC's ethical dilemma injections (\url{https://tinyurl.com/GenCyber5years} [Access date: June 15, 2025]).
\item \textbf{Phase 5: Reflective Exploration:} Many camps conclude with final projects, presentations, and structured reflection activities where students review what they have learned and consider how to apply it (\url{https://tinyurl.com/GenCyberProjects} [Access date: June 15, 2025]). This capstone experience mirrors ECAC's emphasis on reflection to consolidate learning and foster long-term interest.
\end{itemize}
By mapping the well-documented and successful practices of the national GenCyber program onto the ECAC framework, we can see a clear, real-world precedent. This alignment demonstrates that the structure proposed by ECAC is not merely theoretical but is already being applied in successful K-12 cybersecurity education initiatives, thus validating its design and practical potential.

The GenCyber program does not represent a pre-existing version of ECAC. Instead, it serves as powerful evidence from the field that validates ECAC's core tenets. The contribution of the ECAC framework is to synthesize these successful but often disparate practices into an explicit, comprehensive, and theoretically-grounded pedagogical model that can be intentionally replicated, studied, and improved upon.

\section{Implications for Stakeholders and Recommendations}
\label{sec8}
The adoption of the ECAC framework would carry significant implications for various stakeholders in the education and cybersecurity communities. By reframing how and why we teach cybersecurity in K-12, ECAC prompts changes in curriculum planning, teacher training, policy support, and even workforce development strategies. Here we outline these implications and offer recommendations:

\subsection{For Educators and Curriculum Developers}
\label{sec8.1}
Implementing ECAC means that teachers are not just imparting technical content, but also acting as cognitive mentors and ethical role models. Educators would need to embrace a more facilitator-oriented pedagogy. Instead of traditional lectures, they guide students through discovery, ask open-ended questions, and manage collaborative and reflective activities. This requires training and comfort with student-centered learning. Professional development programs must be updated to include both basic cybersecurity concepts and the pedagogical methods of cognitive apprenticeship (e.g. how to effectively model problem-solving or conduct a guided reflection). Curriculum developers, meanwhile, should revisit K-12 learning standards and integrate cybersecurity (with ethics) as a thread across grade levels. ECAC could be used as a blueprint to create grade-appropriate modules that fit into existing subjects – for example, a module on “ethical use of information” in a middle school ICT class, or a series of CTF-like exercises in a high school programming course. One recommendation is to start small: educators can pilot ECAC-inspired activities in a single unit or after-school club, then expand gradually as they gain confidence. Collaboration among teachers is also key – those with more tech experience can support those who are new to cybersecurity. School administrators should recognize the time and effort teachers invest in learning these new skills and possibly provide incentives or release time for training. Ultimately, the implication is that teaching cybersecurity is no longer confined to specialists; with ECAC, it becomes a shared responsibility across the faculty (much like literacy or digital citizenship), and teachers will need support to fulfill this new role.

\subsection{For School Leaders and Education Policymakers}
\label{sec8.2}
At an institutional and policy level, ECAC calls for cybersecurity to be elevated from an extracurricular niche to a core component of 21st-century education. School leaders should consider integrating ECAC elements into the formal curriculum, which might entail curriculum redesign and alignment with standards. Policymakers and curriculum standards bodies might take cues from ECAC to incorporate cybersecurity and ethics outcomes in state or national standards for technology education. A practical implication is the allocation of resources: schools may need funding for technology (like secure lab environments or competition licenses), and policymakers could create grants or initiatives to support this. Additionally, bridging the teacher expertise gap is partly a policy issue – requiring investment in teacher training at scale. Education departments could partner with universities or cybersecurity organizations to develop certification programs or online courses for teachers to get up to speed. Another policy implication is around assessment and accountability: if cybersecurity (with ethics) becomes part of the curriculum, new forms of assessment will be needed to measure success (standardized tests in this domain are nearly non-existent). Policymakers should be cautious not to reduce ECAC to something testable by multiple-choice; instead, support authentic assessment strategies (like portfolios or challenge-based evaluations) that capture the holistic skills ECAC aims to develop. Finally, ensuring equity is a policy responsibility – states or districts implementing ECAC should monitor participation among different student groups and intervene if gaps emerge (for example, by providing extra support to schools serving disadvantaged communities). In summary, ECAC’s implementation would signal a shift in educational priorities, and leaders at the school and policy level must champion this shift by integrating cybersecurity education into mainstream educational mandates and providing the needed infrastructure and training.

\subsection{For Industry and Workforce Development}
\label{sec8.3}
Although ECAC is a K-12 education framework, it ultimately connects to the broader goal of developing a strong, ethical cybersecurity workforce. Industry stakeholders and workforce planners should take note of ECAC’s approach because it promises a pipeline of talent that is not only skilled earlier in life, but also more well-rounded in critical thinking and ethics. Companies often lament the shortage of qualified cybersecurity professionals; ECAC addresses the root of that problem by inspiring and preparing students from a young age. Moreover, by drawing in diverse participants, ECAC can lead to a more diverse workforce down the line, which is known to spur innovation in security solutions. Industry could support ECAC implementation by providing content and context expertise – for instance, professionals could help design realistic scenarios or even participate in mentoring (perhaps virtually joining a class to model an incident response). This kind of school-industry partnership can enrich the authenticity of ECAC activities. From a workforce development perspective, one implication is that if ECAC or similar programs spread, the average baseline knowledge of high school graduates in areas like cybersecurity and digital ethics will rise. This could eventually reduce the training burden on employers and higher education, as incoming employees or college students possess foundational cybersecurity literacy. However, it also means curricula at higher education may need to advance the starting point (since freshmen might already know what used to be taught in intro courses). Industry and post-secondary institutions should stay aligned with K-12 developments to adjust their expectations and programs. A recommendation here is for industry consortia or government workforce programs to invest in K-12 initiatives like ECAC as a long-term talent strategy – sponsoring competitions, offering internships for high schoolers, or donating equipment to schools. The payoff is a generation of cybersecurity-aware individuals, some of whom will pursue cybersecurity careers and all of whom will be more security-conscious citizens, ultimately strengthening the security posture of society.

\subsection{Implementation Challenges and Strategies}
\label{sec8.4}
Translating the ECAC framework from theory into practice will come with challenges. Anticipating these, we outline strategies to handle implementation hurdles:
\begin{itemize}
\item \textbf{Curriculum Integration vs. Time Constraints:} One challenge is fitting ECAC activities into an already packed school schedule. Teachers may worry about sacrificing time from tested subjects to do cybersecurity projects. The strategy here is to integrate with existing learning goals – for example, use a cybersecurity scenario to also teach critical thinking or teamwork skills that are part of the general curricula. Schools can start by replacing or enhancing a portion of computer classes or advisory periods with ECAC content. Showing that ECAC supports broader competencies (problem-solving, digital literacy) can justify the time investment. Gradual adoption is key; perhaps one competition module per semester, then increasing frequency as it proves its value.
\item \textbf{Teacher Training and Support:} As mentioned, not all teachers currently have the background to lead cybersecurity competitions. Beyond formal professional development, creating a support network is vital. This can include online forums for ECAC educators to share experiences, a repository of ready-made lesson plans and challenges, and access to mentors (maybe local cybersecurity experts willing to volunteer). Providing teachers with co-teaching opportunities or classroom visits by specialists during initial implementation can build confidence. Importantly, teacher prep should include not just technical training but also how to facilitate ethical discussions – many teachers are not used to guiding conversations on moral dilemmas, so giving them case studies and practice in this is important.
\item \textbf{Resource and Infrastructure Setup:} For schools with limited IT infrastructure, setting up even a simple CTF environment might be daunting. To mitigate this, ECAC recommends using cloud-based platforms or “CTF-in-a-box” solutions where possible. These are hosted environments that teachers can launch without needing internal servers. Additionally, where internet access is an issue, offline or local network versions of challenges can be provided (some research has already explored “CTF unplugged” models for offline play). Partnering with organizations like universities can also help – they often have resources and platforms that schools can leverage. A phased approach to resources means starting with what’s readily available (e.g., using a free public CTF challenge from \textit{PicoCTF} for an assignment) and then building a dedicated infrastructure if the program expands.
\item \textbf{Ensuring Safety and Legal/Ethical Boundaries:} When teaching real cybersecurity techniques, even in a controlled way, there’s a risk that students might misuse skills or inadvertently break rules (e.g., probing real websites). Implementation guidelines must include clear acceptable use policies and supervision measures. Schools should have students sign honor codes and educate them early on about what is off-limits (for instance, they can practice hacking on provided simulations but not on the school network or public sites). Technical controls can sandbox student activities. This challenge is best met by transparency and education – making ethical use of skills the very fabric of ECAC (which it is) and involving parents and the school community so they understand the goals and safeguards. Involving an ethics or legal studies teacher when devising dilemmas could also strengthen this aspect.
\item \textbf{Assessment and Evaluation:} Traditional grading may not directly apply to an ECAC activity. Educators might wonder how to grade a competition or a team-based project. To address this, ECAC recommends authentic assessment methods: rubrics that credit not just “flag capture” but teamwork, reasoning process, and reflection quality. For example, a portion of the grade could come from a student’s reflection essay or a presentation after the competition, rather than the competition results alone. This ensures students take the reflection seriously and don’t feel all is lost if they didn’t solve many flags – they can demonstrate learning in other ways. Over time, collecting data on student outcomes (both quantitative, like improvement in challenge completion, and qualitative, like growth in ethical reasoning as seen in their writings) will be important to evaluate ECAC’s effectiveness and convince stakeholders of its value.
\end{itemize}
Despite these challenges, the overall recommendation is to approach ECAC implementation iteratively and supportively. Small pilot programs can iron out kinks, and successful strategies can be documented and shared. Institutions should be patient and allow the framework to mature – it may take a few cycles for teachers to fully adjust to the cognitive apprenticeship style and for students to adapt to the idea that a “competition” involves writing and discussions. With commitment and careful management, the ECAC framework can be adopted in diverse school settings, bringing its promised benefits to fruition.

\section{Limitations and Future Research}
\label{sec9}
While the ECAC framework is built on solid theoretical ground and addresses many practical concerns, it is not without limitations. First, ECAC is conceptualized for now – it has been validated here through theory, scenarios, and a use case, but not yet proven in a wide-scale empirical study. As such, one limitation is the lack of direct evidence on its efficacy compared to other methods. Classrooms and student populations vary greatly, and what works in one context may need adjustment in another. Future research is needed to empirically test and refine ECAC. Key avenues for research include:
\begin{itemize}
\item \textbf{Longitudinal Impact Studies:} Following students who learn via ECAC over several years would provide insight into how it shapes their skills and attitudes long-term. For example, do elementary students who start with ECAC-inspired activities show greater cybersecurity interest and literacy by high school? Are ECAC alumni more likely to pursue computing or exhibit stronger ethical reasoning than peers? Such studies should measure not only technical competencies (perhaps via standardized tasks or CTF performance) but also cognitive gains (problem-solving tests) and moral development (ethical dilemma responses). A longitudinal approach can validate whether the staged, cumulative nature of ECAC truly yields compounding benefits over a K-12 span.
\item \textbf{Comparative Effectiveness Research:} It is important to compare ECAC with existing pedagogies. Controlled experiments or quasi-experiments could be set up where, for instance, one group of schools uses a traditional CTF club program and another uses ECAC-integrated curriculum, to see differences in outcomes. Even shorter-term studies, like implementing ECAC for one semester and comparing student engagement and learning to a previous semester that used a different approach, would be valuable. These comparisons will help pinpoint which components of ECAC are most impactful and whether the added complexity of integrating ethics and cognitive apprenticeship indeed produces statistically and educationally significant improvements.
\item \textbf{Teacher Professional Development Models:} Since teacher readiness is a pivotal factor, research should examine how best to train and support teachers for ECAC. This might involve pilot programs that train a cohort of teachers and track their confidence, teaching practices, and student results. By experimenting with different formats (workshops, mentorship pairings, online courses), the education community can learn what enables teachers to effectively become cognitive coaches and ethical mentors in cybersecurity. Feedback from teachers can also guide simplifications or enhancements to the framework to make it more teacher-friendly.
\item \textbf{Inclusivity and Adaptation Studies:} Another research direction is exploring how ECAC can be optimized for diverse populations. Does the framework need modifications for students with special educational needs or disabilities? How does it play out in different cultural contexts or countries? Ensuring the framework is inclusive might involve tailoring scenarios to be culturally relevant or addressing language barriers in technical content. Research could also specifically measure ECAC’s impact on underrepresented groups’ participation and interest in cybersecurity, validating the claim that it broadens engagement. Any gaps identified (e.g., if some group still lags) would direct further adaptations.
\item \textbf{Technology and Tool Development:} Implementing ECAC at scale might benefit from dedicated tools – for example, platforms that integrate challenge solving with reflection journals and teacher dashboards. Research and development can focus on technological supports that embody ECAC principles: perhaps an online system that not only hosts CTF challenges but also pops up ethical dilemma prompts at set intervals, and allows teachers to monitor both challenge progress and student reflections in one place. User studies on such tools in classrooms would guide their design. Additionally, evaluating existing CTF platforms for their suitability in ECAC (and modifying them as needed) is a practical area of inquiry.
\end{itemize}
In acknowledging these limitations and outlining future research, we underscore that \textbf{ECAC is a framework in its infancy}. It presents a compelling approach, but it will require iterative refinement. Pilots and studies will likely lead to adjustments – for instance, perhaps certain phases need more emphasis, or teachers might report that five phases are too many and they prefer condensing to four. The framework should be seen as dynamic, subject to continuous improvement as evidence comes in. Crucially, any future research should also keep an eye on unintended consequences: for example, could making a competition part of class grades dampen the fun for some students? Does heavy emphasis on ethics ever reduce some students’ competitive drive in a way that affects motivation? Such questions would be healthy to explore to ensure ECAC finds the right balance in practice.

Despite the need for further validation, the rationale for ECAC is strong. The limitations present research opportunities, and the authors encourage the academic and educational community to treat ECAC as a hypothesis to test, rather than a finished product – one that, if supported, could significantly advance how we teach and inspire the next generation in cybersecurity.

\section{Conclusion}
\label{sec10}
“\textbf{Beyond the flag}” – the guiding vision of this work – encapsulates the shift from viewing cybersecurity competitions as mere games of capture-and-score, to harnessing them as vehicles for rich education and character development. The Ethical-Cognitive Apprenticeship in Cybersecurity framework offers a novel, comprehensive approach to integrating cybersecurity competitions into K-12 education. By systematically blending the excitement of CTF-style challenges with the proven pedagogical methods of cognitive apprenticeship and an unwavering emphasis on ethical thinking, ECAC transforms a fun extracurricular activity into a meaningful learning journey. It moves the focus from winning flags to learning from the journey: students progressively acquire technical skills, learn to think like experts, collaborate inclusively, and crucially, internalize the ethical responsibilities that come with cybersecurity knowledge.

The analysis presented in this section suggests that ECAC is theoretically sound and pedagogically robust. It directly addresses current gaps in K-12 cybersecurity education by providing structure, context, and inclusion where often there is randomness or exclusivity. If implemented, ECAC has the potential to produce not only more skilled young people – capable of tackling cyber challenges – but also more conscientious digital citizens who approach technology with a critical and ethical lens. In a world where cyber threats are ever-evolving and the line between digital play and real-world consequences is increasingly thin, such an education is timely and essential.

Ultimately, the success of ECAC or any similar initiative will depend on collective will – of educators willing to innovate, of institutions investing in new curriculum, and of communities recognizing that cybersecurity savvy is as important as reading or math in the modern era. The hope is that this framework provides a blueprint and an inspiration for stakeholders to act. By preparing students through an apprenticeship of both mind and character, we equip them not just to navigate the digital world, but to positively shape it. The flag captured at the end of an ECAC challenge is not just a puzzle solved; it is a symbol of a learner empowered with knowledge, guided by ethics, and ready for the challenges ahead.

\section*{Acknowledgements}
We sincerely thank the anonymous reviewers and editors for their valuable comments and constructive feedback, which greatly contributed to improving the quality and clarity of this article.

\section*{Disclosure statement}

No potential conflict of interest was reported by the authors.

\section*{Declaration of generative AI in scientific writing}

During the preparation of this work, the authors used Chatgpt-4o to rewrite or paraphrase the written paragraphs
in an academic tone. After using this tool, the authors reviewed and edited the content as needed and take full
responsibility for the content of the publication.

\appendix
\section{Appendix}
\label{app1}

A list of the 25 studies used in this research is available here: \url{https://tinyurl.com/StudiesInResearch}

\bibliographystyle{ACM-Reference-Format}
\bibliography{Reference}

\end{document}